\documentclass[pra,aps,twocolumn,superscriptaddress,showpacs,nofootinbib,floatfix]{revtex4}

\usepackage{amsmath,amssymb,amsbsy,graphicx,subfigure,bm}

\newcommand{\eq}[1]{Eq.~(\ref{#1})}
\newcommand{\ineq}[1]{Ineq.~(\ref{#1})}
\newcommand{\ket}[1]{|#1\rangle}
\newcommand{\bra}[1]{\langle#1|}

\begin{document}

\title{Measurement-induced disturbances and nonclassical correlations of Gaussian states}

\date{March 15, 2011}

\author{Ladislav Mi\v{s}ta, Jr.}
\affiliation{Department of Optics, Palack\' y University, 17.~listopadu 12,  771~46 Olomouc, Czech Republic}
\affiliation{School of Physics and Astronomy, University of St.~Andrews, North Haugh, St.~Andrews, Fife, KY16 9SS, Scotland}
\author{Richard Tatham}
\affiliation{School of Physics and Astronomy, University of St.~Andrews, North Haugh, St.~Andrews, Fife, KY16 9SS, Scotland}
\author{Davide Girolami}
\affiliation{School of Mathematical Sciences, University of Nottingham, University Park, Nottingham NG7 2RD, United Kingdom}
\author{Natalia Korolkova}
\affiliation{School of Physics and Astronomy, University of St.~Andrews, North Haugh, St.~Andrews, Fife, KY16 9SS, Scotland}
\author{Gerardo Adesso}
\affiliation{School of Mathematical Sciences, University of Nottingham, University Park, Nottingham NG7 2RD, United Kingdom}

\begin{abstract}
We study quantum correlations beyond entanglement in two--mode Gaussian states of continuous variable systems, by means of the  measurement-induced disturbance (MID) and its ameliorated version (AMID). In analogy with the recent studies of the Gaussian quantum discord, we define a Gaussian AMID by constraining the optimization to all bi-local Gaussian positive operator valued measurements. We solve the optimization explicitly for relevant families of states, including squeezed thermal states. Remarkably, we find that there is a finite subset of two--mode Gaussian states, comprising pure states, where non-Gaussian measurements such as photon counting are globally optimal for the AMID and realize a strictly smaller state disturbance compared to the best Gaussian measurements. However, for the majority of two--mode Gaussian states the unoptimized MID provides a loose overestimation of the actual content of quantum correlations, as evidenced by its comparison with Gaussian discord. This feature displays strong similarity with the case of two qubits. Upper and lower bounds for the Gaussian AMID at fixed Gaussian discord are identified. We further present a comparison between Gaussian AMID and Gaussian entanglement of formation, and classify families of two--mode states in terms of their Gaussian AMID, Gaussian discord, and Gaussian entanglement of formation. Our findings provide a further confirmation of the genuinely quantum nature of general Gaussian states, yet they reveal that non-Gaussian measurements can play a crucial role for the optimized extraction and potential exploitation of classical and nonclassical correlations in Gaussian states.
\end{abstract}


\maketitle

\section{Introduction}\label{secIntro}

One of the seminal findings triggering the development of quantum information theory is that there exist nonlocal
correlations among subparts of quantum systems that do not emerge in a classical scenario. These nonclassical
correlations, commonly identified with entanglement, can be exploited to manipulate and transmit information in novel and enhanced ways
\cite{nielsen} going beyond the possibilities of classical physics. Consequently, an increasing interest in their study has risen in recent years \cite{hororev}.

Interestingly, signatures of correlations having no classical counterpart can be traced even in separable
(non-entangled) states, but their nature is rather different from entanglement \cite{OZ,HV}. In fact, while
entanglement is a consequence of the superposition principle, more general forms of nonclassical correlations
arise essentially from the noncommutativity of quantum observables.
%Not only are \emph{truly} classical
%states just a subset of the separable states, but it is
When speaking about composite systems, separable states are often perceived as essentially classical. However, \emph{truly} classical states, i.e. states which contain only classical correlations, represent just a subset of separable states. Moreover, it is
also possible to show that almost all  separable states possess a finite amount of nonclassical correlations
\cite{ferraro}. This has fueled a still unsettled debate, and an active stream of research, to decide whether
separable states containing nonclassical correlations can be also directly useful for quantum information tasks
\cite{piani,dattaqc,barbieri,brasilearxiv}.

To quantitatively assess various aspects of entanglement, several entanglement measures have been adopted and extensively studied \cite{hororev}. In a similar fashion, more
recently a zoology of indicators of nonclassical correlations (in separable or entangled states) have been introduced
\cite{OZ,HV,MID,piani,moelmer,amid,misure}, among which the most popular being {\it quantum discord} \cite{OZ,HV}.
A nice facet of discord is that it has an immediate characterization in information theory, it is endowed with
operational interpretations \cite{operdiscord}, and its evaluation for bipartite states of simple quantum systems
such as two-level systems (qubits) is conceptually straightforward, albeit technically hard.

 Studies on nonclassicality indicators are not restricted to finite dimensions, but have also been extended to quantum
 systems with infinite-dimensional state space,  where correlations are encoded between continuous variables (CVs), that is, variables with continuous  spectra. They are represented by, e.g., modes of the electromagnetic field described
 by quadrature amplitudes. For these systems,
 a privileged role is played by the states possessing a Gaussian Wigner function, the so called Gaussian states,
 as they are easy to handle both theoretically \cite{ourreview} and experimentally \cite{gaussexp}. Although Gaussian states are
 sometimes flagged as the ``most classical'' class of continuous-variable (CV) states because of
 the positivity of their Wigner function, recent results of the nonclassicality indicators show the opposite. Namely,
 a recently derived analytical form of quantum discord for two--mode Gaussian states \cite{AD_10,Giorda_10} reveals
 that, contrarily to the above naive categorization, all non-product bipartite Gaussian states have a nonzero discord
 and so exhibit nonclassical correlations.

Another nonclassicality indicator frequently employed in the literature is the \emph{measurement-induced disturbance} or \emph{MID},
introduced in \cite{MID}. A good property of  MID is the intrinsic symmetry under swapping of the subsystems
(unlike discord) but the main flaw is that it does not incorporate any optimization over local measurements, therefore usually returning an an overestimation of the actual amount of nonclassical correlations. The necessity of a more
faithful nonclassicality quantifier motivated the introduction of a new indicator called \emph{AMID}
({\it ameliorated measurement-induced disturbance}) which is an ameliorated synthesis of the discord and MID, being both symmetric as
MID and optimized as discord. It has been applied to characterize nonclassical correlations of arbitrary two--qubit mixed states
\cite{amid}. The AMID is defined as the difference between the total (quantum) and the classical mutual information \cite{terhal,moelmer,amid,brasilesymm}, where the latter quantity represents the maximum classical correlations
that can be extracted from a bipartite state via local measurements on subsystems. Here an optimization over
all possible bi-local measurements is involved, which represents the nontrivial part in calculation of the AMID.

In the present work, we continue the programme of characterizing quantum versus classical correlations beyond entanglement in CV systems. Specifically, we extend the definitions of  MID and AMID to the paradigmatic class of two--mode Gaussian states. Contrary to the case of discord, which has been computed for Gaussian states restricting the optimization to Gaussian measurements only \cite{Giorda_10,AD_10}, for the nonclassicality indicators considered here non-Gaussian measurements play an important role. The MID is in fact obtained as the state disturbance induced by local projections onto Fock states (photon counting), a clearly non-Gaussian measurement. This can be done analytically for simple cases (such as pure two--mode squeezed states) and requires numerical evaluation in more general Gaussian states. For the AMID, a competition between photon counting and optimized Gaussian measurements takes place in the maximization of the classical mutual information. We define a Gaussian version of AMID (Gaussian AMID), restricted to Gaussian measurements in analogy with the Gaussian discord \cite{Giorda_10,AD_10}, and provide the analytical framework for its computation, obtaining simple closed forms for relevant families of states, including pure states and squeezed thermal states. We compare this measure with MID and Gaussian discord, elucidating similarities and differences with the two--qubit case \cite{amid}. Similarly, we find that also for Gaussian states, in most cases, MID  significantly overestimates quantum correlations, and we provide evidence for states with nearly vanishing discord but arbitrarily large MID. Yet, quite interestingly, we find a finite region of two--mode Gaussian states where MID is strictly smaller than the optimal Gaussian AMID, meaning that in those instances non-Gaussian measurements are necessarily optimal for the calculation of the classical mutual information and for the AMID. This subset of states includes, surprisingly, pure two--mode squeezed states, for which local photon countings are found to be therefore ``less disturbing'' than local homodyne detections (which realize the optimal Gaussian measurements in this case), and allow one to extract strictly more strongly correlated classical random variables from the pure Gaussian states. The gap  between photon counting and homodyne detection in the degrees of classical mutual information and AMID persists even in the limit of infinite squeezing.
Such a finding is quite unexpected and certainly deserves further investigation with an eye on potential practical applications. Yet, this agrees in spirit with  with a series of somehow related results: For example,  the best operation to clone Gaussian coherent states, as well as the best partial measurement achieving the optimal information/disturbance tradeoff for the same type of state, and the best measurement localizing maximum
bipartite entanglement for Gaussian states, are all known to be non-Gaussian \cite{Cerf05,Mista06,Fiurasek_07}. Recall also that non-Gaussian measurements are necessary for universal CV quantum computation with Gaussian cluster states \cite{menicucci}.

The paper is organized as follows. In Section \ref{secPrelim} we set up the notation and recall the main concepts of Gaussian states and Gaussian measurements, followed by an overview of the nonclassicality measures landscape. The derivation of a manageable expression of MID (associated to local Fock projections) for two--mode Gaussian states, in closed form for pure states, is presented in Section \ref{secMID}.
%We remark the  non--Gaussian nature of the local measurements and, comparing the MID with the discord, the very close similarity to the two--qubit case evaluated in \cite{amid}: In both cases, MID is an upper bound of discord dramatically overestimating quantum correlations.\\
In Section \ref{secGAMID} we face the evaluation of the AMID for arbitrary two--mode Gaussian states. We bound AMID from above by the minimum between the MID and the Gaussian AMID. The latter is defined by restricting the optimization to bi-local Gaussian measurements, adopting the techniques used for discord in \cite{Giorda_10,AD_10}, and we provide closed analytical forms for it on special families of two--mode Gaussian states.
Section \ref{secDiscuss} presents a thorough comparison between MID, Gaussian AMID and Gaussian quantum discord on random two--mode Gaussian states, highlighting hierarchical and ordering relations between the three measures. Upper and lower bounds to the Gaussian  AMID at fixed Gaussian discord are identified. The states where non-Gaussian measurements are necessary to reach the minimum in the AMID are amply discussed and characterized. In Section \ref{secEnt}, we compare the Gaussian AMID with the Gaussian entanglement of formation (Gaussian EoF) \cite{giedke03,geof}, showing on the basis of numerical evidence that the Gaussian AMID is always greater or equal than the Gaussian EoF for all two--mode Gaussian states, and admits also an upper bound at fixed Gaussian EoF (similarly to what observed for discord \cite{AD_10}). A visual classification of the special class of symmetric squeezed thermal states based on their degrees of Gaussian EoF, Gaussian discord and Gaussian AMID is also provided.
Finally, Section \ref{secConcl} summarizes the results we obtained, underlining the main implications and delivering hints on possible future applications.

\section{Preliminaries}\label{secPrelim}

\subsection{Gaussian states and measurements}\label{subsecGauss}
We consider two modes $A$ and $B$ described by the vector $\hat{\xi}=(\hat{x}_{A},\hat{p}_{A},\hat{x}_{B},\hat{p}_{B})^{T}$
of quadrature operators $\hat{x}_{j},\hat{p}_{j}$, $j=A,B$, satisfying the canonical commutation rules
that can be expressed as $[\hat{\xi}_{j},\hat{\xi}_{k}]=i\Omega_{jk}$, where
%%%%%%%%%%%%%%%%%%%%%%%%%%%%%%%%%%%%%%%%%%%%%%%%%%%%%%%%%%%%%%%%%%%%%%%%%%%%%%%%%%%%%%%%%%%%%%%
\begin{eqnarray}\label{Omega}
\Omega=J\oplus J,\quad J=\left(\begin{array}{cc}
0 & 1 \\
-1 & 0 \\
\end{array}\right).
\end{eqnarray}
%%%%%%%%%%%%%%%%%%%%%%%%%%%%%%%%%%%%%%%%%%%%%%%%%%%%%%%%%%%%%%%%%%%%%%%%%%%%%%%%%%%%%%%%%%%%%%%%
A two--mode Gaussian state $\hat{\rho}_{AB}$ can be represented in phase space by a Gaussian Wigner function
%%%%%%%%%%%%%%%%%%%%%%%%%%%%%%%%%%%%%%%%%%%%%%%%%%%%%%%%%%%%%%%%%%%%%%%%%%%%%%%%%%%%%%%%%%%%%%%%
\begin{eqnarray}\label{W}
W(r)=\frac{1}{\pi^2\sqrt{\det\gamma}}e^{-(r-d)^{T}\gamma^{-1}(r-d)},
\end{eqnarray}
%%%%%%%%%%%%%%%%%%%%%%%%%%%%%%%%%%%%%%%%%%%%%%%%%%%%%%%%%%%%%%%%%%%%%%%%%%%%%%%%%%%%%%%%%%%%%%%%
where $r=(x_{A},p_{A},x_{B},p_{B})^{T}$ is the radius vector in phase space, $d=\mbox{Tr}(\hat{\rho}_{AB}\hat{\xi})$
is the vector of phase-space displacements and $\gamma$ is the covariance matrix (CM) with elements $\gamma_{jk}=2\mbox{Re}\mbox{Tr}[\hat{\rho}_{AB}(\hat{\xi}_{j}-d_{j})(\hat{\xi}_{k}-d_{k})]$,
$j,k=1,\ldots,4$. Any CM $\gamma$ has to satisfy the constraints $\gamma > 0$ and $\gamma + i \Omega \ge 0$
(positivity and uncertainty principle) to ensure that it is associated to a physical Gaussian state.

The CM contains a complete information about the correlations in a given Gaussian state \cite{ourreview}.
By means of local symplectic (unitary on the Hilbert space) operations, which leave
correlations and entropic quantities invariant, the CM of a two--mode Gaussian state can always
be reduced to a simple standard form
\begin{equation}\label{gamma}
\gamma=\left( \begin{array}{cc}
A  & C \\
C^T & B
\end{array}
\right) = \left(
\begin{array}{cccc}
 a & 0 & c_1 & 0 \\
 0 & a & 0 & c_2 \\
 c_1 & 0 & b & 0 \\
 0 & c_2 & 0 & b
\end{array}
\right),
\end{equation}
where we can assume $c_1 \ge |c_2| \ge 0$ without any loss of generality. States with $a=b$ are said to
be symmetric, while states with $c_2=\pm c_1$ constitute the important family of {\it squeezed thermal states}.
Pure two--mode Gaussian states are special instances of symmetric squeezed thermal states, with zero temperature
($\det\gamma=1$), i.e., they are locally equivalent to two--mode squeezed vacuum states,
with standard form covariances
\begin{equation}\label{puri}
a=b=\cosh(2r)\,,\quad c_1=-c_2=\sinh(2r)\
\end{equation}
with $r$ being the squeezing parameter.

Gaussian states can be produced, manipulated and detected in the laboratory with a high degree of control \cite{gaussexp}. Some measurements such as photon counting turn a Gaussian state into a non-Gaussian one. On the other hand, Gaussian measurements play a special role as being those that  map Gaussian states into Gaussian states. These measurements coincide with the standard toolbox of linear optics, i.e.,~can be realized using beam splitters, phase shifters, squeezers, appending auxiliary vacuum states, and performing balanced homodyne detection (BHD).
Any such a measurement is described by a positive operator valued measure (POVM) of the form \cite{Fiurasek_07}
\begin{equation}\label{POVM}
\hat{\Pi}^G_j(d_j)= \frac{1}{2\pi} \hat{D}_j(d_j)\hat{\Pi}^{G}_j \hat{D}_j^\dagger(d_j),\quad j=A,B.
\end{equation}
Here the seed element $\hat{\Pi}^{G}_j$ is a normalized density matrix of a generally mixed single--mode Gaussian state
with CM $\gamma_j$ and zero displacements, $\hat{D}_j(d_j)=\exp(-id_{j}^{T}J\hat{\xi_{j}})$ stands for the
displacement operator, where $\hat{\xi}_{j}=(\hat{x}_{j},\hat{p}_{j})^{T}$ and
$d_j=(d_{j}^{(x)},d_{j}^{(p)})^{T}$ is a vector of certain linear combinations of the measurement outcomes of BHDs.
The POVM (\ref{POVM}) satisfies the completeness condition
%%%%%%%%%%%%%%%%%%%%%%%%%%%%%%%%%%%%%%%%%%%%%%%%%%%%%%%%%%%%%%%%%%%%%%%%%%%%%%%%%%%%%%%%%%%%%%%%%%%%%%%
\begin{equation}
\frac{1}{2\pi} \int \hat{D}_j(d_j) \hat{\Pi}^{G}_j \hat{D}_j^\dagger(d_j){\rm d}^{2}d_{j}= \hat{\openone}_j,
\label{completness}
\end{equation}
%%%%%%%%%%%%%%%%%%%%%%%%%%%%%%%%%%%%%%%%%%%%%%%%%%%%%%%%%%%%%%%%%%%%%%%%%%%%%%%%%%%%%%%%%%%%%%%%%%%%%%%
where
${\rm d}^{2}d_{j}={\rm d}d_{j}^{(x)}{\rm d}d_{j}^{(p)}$, 
following from Schur's lemma \cite{D'Ariano_04} and the normalization condition $\mathrm{Tr}[\hat{\Pi}^{G}_j]=1$.

\subsection{Measures of quantum correlations}\label{subsecMeas}

Here we briefly review some of the most prominent measures of quantum correlations beyond entanglement, recently proposed to identify the genuinely nonclassical portion of the total correlations in generally mixed states of bipartite quantum systems. Achieving a proper understanding of the structure and nature of correlations in relevant systems is important for gaining insights into foundational aspects of quantum theory, and is particularly imperative in view of practical applications, as quantum correlations yield the key resources to overcome classical systems in quantum information protocols. For completeness, we also recall the definition of the entanglement of formation.

{\it Quantum mutual information.}---The total amount of (classical and quantum) correlations in the state of a bipartite quantum system can be reliably quantified in terms of the quantum mutual information \cite{Stratonovich_65,Adami_97}
\begin{equation}\label{MI}
{\cal I}_q(\hat{\rho}_{AB})={\cal S}(\hat{\rho}_{A})+{\cal S}(\hat{\rho}_{B})-{\cal S}(\hat{\rho}_{AB})\,,
\end{equation}
where ${\cal S}(\hat{\rho})=-\mbox{Tr}(\hat{\rho}\ln\hat{\rho})$ is the von Neumann entropy and $\hat{\rho}_{A,B}$ are the reduced states of subsystems $A$ and $B$, respectively.
The quantum mutual information ${\cal I}_{q}(\hat{\rho})$ of a generic Gaussian state $\hat{\rho}$
can be easily calculated using the formula for the von Neumann entropy of an $N$-mode Gaussian state $\hat{\rho}$ \cite{Holevo_01},
${\cal S}(\hat{\rho})=\sum_{i=1}^{N}\mathfrak{F}(\nu_{i})$, where $\nu_{i}$ are the symplectic eigenvalues \cite{Williamson_36}
of the CM of the state and \begin{equation}\label{effe}
\mathfrak{F}(x)=\left(\frac{x+1}{2}\right)\ln\left(\frac{x+1}{2}\right)-\left(\frac{x-1}{2}\right)\ln\left(\frac{x-1}{2}\right).
\end{equation}
For two--mode Gaussian states $\hat{\rho}_{AB}$, the two global symplectic eigenvalues $\nu_{\pm}$ are defined by $2\nu_{\pm}^2 = \Delta \pm \sqrt{\Delta^2-4\det\gamma}$, with $\Delta = \det A+\det B+2\det C$, see \eq{gamma} \cite{extremal}. The mutual information of a two--mode Gaussian state with CM $\gamma$ as in \eq{gamma} is thus
\begin{equation}\label{MIG}
{\cal I}_q(\hat{\rho}_{AB}) = \mathfrak{F}\left(\sqrt{\det A}\right)+\mathfrak{F}\left(\sqrt{\det B}\right) - \mathfrak{F}\left(\nu_+\right)- \mathfrak{F}\left(\nu_-\right)\,,
\end{equation}
with $\mathfrak{F}(x)$ defined in \eq{effe}.

{\it One-way classical correlations and quantum discord.}--- For any bipartite state whose correlations are purely classical, the mutual information can be equivalently expressed in two alternative forms
\begin{equation}\label{J}
\begin{split}
{\cal J}^\leftarrow(\hat{\rho}_{AB}) &= {\cal S}(\hat{\rho}_{A}) - \inf_{\{\hat\Pi_i\}} {\cal H}_{\{\hat\Pi_i\}}(A|B)\,, \\
{\cal J}^\rightarrow(\hat{\rho}_{AB}) &= {\cal S}(\hat{\rho}_{B}) - \inf_{\{\hat\Pi_i\}} {\cal H}_{\{\hat\Pi_i\}}(B|A)\,,
\end{split}
\end{equation}
with ${\cal H}_{\{\hat\Pi_i\}}(A|B){\equiv}\sum_{i}p_i{\cal S}(\hat{\rho}^i_{A|B})$ being the quantum conditional entropy associated with  the post-measurement density matrix $\hat{\rho}^i_{A|B} = \mbox{Tr}_B[\hat\Pi_i\hat{\rho}_{AB}]/p_i$, obtained upon performing the POVM $\{\hat\Pi_i\}$ on system $B$ ($p_i{=}\mbox{Tr}[\hat\Pi_i\hat{\rho}_{AB}]$); the optimization over the POVMs is necessary to single out the least disturbing measurement to be performed on one subsystem, so that the change of entropy on the other subsystem yields a quantifier of the correlations between the two parts.

For arbitrary bipartite quantum states $\hat{\rho}_{AB}$, including all entangled states and almost all separable states as well \cite{ferraro}, the three quantities in Eqs.~(\ref{MI}) and (\ref{J}) evaluate to different results in general, with ${\cal I}_q \ge {\cal J}^\leftarrow,{\cal J}^\rightarrow$, and the ${\cal J}$ quantities can be interpreted as `one-way classical correlation' measures \cite{HV}. Such a  discrepancy is now recognized as a signature of nonclassicality of the correlations in a given state, and the difference between total correlation [\eq{MI}] and one-way classical correlation [\eq{J}] defines what Ollivier and Zurek baptized as the `quantum discord' \cite{OZ},
\begin{eqnarray}\label{discord}
\mathcal{D}^\leftarrow(\hat{\rho}_{AB}) &=& {\cal I}_q(\hat{\rho}_{AB})-\mathcal{J}^\leftarrow(\hat{\rho}_{AB})\\  &=& {\cal S}(\hat{\rho}_B)-{\cal S}(\hat{\rho}_{AB})+\inf_{\{\hat\Pi_i\}} {\cal H}_{\{\hat\Pi_i\}}(A|B)\,; \nonumber \\
\mathcal{D}^\rightarrow(\hat{\rho}_{AB}) &=& {\cal I}_q(\hat{\rho}_{AB})-\mathcal{J}^\rightarrow(\hat{\rho}_{AB})\\  &=& {\cal S}(\hat{\rho}_A)-{\cal S}(\hat{\rho}_{AB})+\inf_{\{\hat\Pi_i\}} {\cal H}_{\{\hat\Pi_i\}}(B|A)\,. \nonumber
\end{eqnarray}
Quantum discord is an asymmetric measure of quantum correlations which has recently received its first operational interpretations in terms of the quantum state merging protocol \cite{operdiscord} and has spurred a great body of research triggered by the investigation of its potential role as the key resource for mixed-state quantum computation \cite{dattaqc,barbieri,brasilearxiv,ferraro,spinchains,decoherence}.
The presence of the optimization over local measurements in the definition of discord makes its analytical evaluation very hard for general bipartite states. No closed formulas are known  for ${\cal D}$ on arbitrary two--qubit mixed states, other than special cases \cite{luoalber}. Very recently, a Gaussian version of quantum discord has been defined \cite{Giorda_10,AD_10}, where the optimization is restricted to Gaussian POVMs of the type \eq{POVM}, and its closed expression has been derived for arbitrary two--mode Gaussian states \cite{AD_10}. As a consequence of that analysis, it was established that the only classically-correlated Gaussian states are product states, which are completely uncorrelated. However, in the limit of diverging mean energy there can exist two--mode Gaussian states that asymptotically approach so-called classical-quantum states \cite{piani}, where ${\cal D}^\leftarrow \rightarrow 0$ while ${\cal D}^\rightarrow > 0$ \cite{AD_10}. A symmetrized version of quantum discord --- or `two-way discord' --- can be defined as
\begin{equation}\label{twoway}
\mathcal{D}^\leftrightarrow(\hat{\rho}_{AB})=\max\{\mathcal{D}^\leftarrow(\hat{\rho}_{AB}),\mathcal{D}^\rightarrow(\hat{\rho}_{AB})\}\,,
\end{equation}
and in this form becomes vanishing if and only if a state is purely classically correlated \cite{amid}. Throughout the paper, ${\cal D}$ will in general denote the Gaussian quantum discord \cite{Giorda_10,AD_10}, unless explicitly stated.

{\it Measurement-induced disturbance.}--- In order to overcome the difficulties involved in the evaluation of quantum discord, Luo introduced the `measurement-induced disturbance' (MID) as an alternative nonclassicality indicator for bipartite quantum states \cite{MID}. MID is motivated by  the observation that in classical systems, local measurements do not induce disturbance. In particular, a bipartite state containing no quantum correlations is left invariant by the action of any bi-local complete  measurement. On the other hand, even when a state $\hat{\rho}_{AB}$ is a priori nonclassical, any complete bi-local measurement makes it classical as a result of a decoherence-by-measurement process~\cite{MID}. MID is thus defined by restricting the attention to the bi-local complete projective measurement $\hat{\cal E}_{A}\otimes\hat{\cal E}_{B}$ determined by the eigen-projectors $\hat{\cal E}_{j}(k)$ of the marginal states $\hat{\rho}_{j}=\sum_{k}\lambda_{k}\hat{\cal E}_{j}(k)$~($j=A,B$), where $\lambda_{k}$ are corresponding eigenvalues, and reads \cite{MID}
\begin{equation}\label{MID}
{\cal M}(\hat{\rho}_{AB})={\cal I}_q(\hat{\rho}_{AB})-{\cal I}_q[\hat{\cal E}(\hat{\rho}_{AB})]\,,
\end{equation}
where
\begin{eqnarray}\label{postmeasurement}
\hat{\cal E}(\hat{\rho}_{AB})=\sum_{k,l}p_{AB}(k,l)\hat{\cal E}_{A}(k)\otimes\hat{\cal E}_{B}(l)
\end{eqnarray}
is the post-measurement state after local measurements $\hat{\cal E}_{A}$ and $\hat{\cal E}_{B}$ and
$p_{AB}(k,l)=\mbox{Tr}[\hat{\rho}_{AB}\hat{\cal E}_{A}(k)\otimes\hat{\cal E}_{B}(l)]$ is the probability
of obtaining the outcome $(kl)$. The post-measurement state is
obviously fully classical which implies that its quantum mutual information (\ref{MI}) coincides with the classical
mutual information of the distribution $p_{AB}$ given by ${\cal I}(A:B)={\cal H}(p_A)+{\cal H}(p_B)-{\cal H}(p_{AB})$
\cite{Shannon_48} with ${\cal H}$ being the Shannon entropy, where $p_{A}$ and $p_{B}$ are
reduced probability distributions of the distribution $p_{AB}$. Hence we can rephrase MID as
\begin{equation}\label{MID2}
{\cal M}(\hat{\rho}_{AB})={\cal I}_q(\hat{\rho}_{AB})-{\cal I}(A:B)\,.
\end{equation}

The MID  quantifies the quantumness of correlations in terms of the state disturbance after local measurements, but with important differences compared to quantum discord: (i) both subsystems are locally probed; (ii) there is no optimization over the local measurements, which are chosen to be the marginal eigen-projectors for every quantum state \footnote{Notice also that this choice of measurements makes MID not uniquely defined on bipartite states whose reduced density matrices have a degenerate spectrum \cite{moelmer}, as it is the case for states with maximally mixed marginals.}.
Although such a quantity is easily computable in arbitrary-dimensional systems, and has found widespread applications in several investigations \cite{useMID}, a number of studies have pointed out that MID is clearly an unfaithful and non-refined measure of the nonclassicality of correlations in bipartite states \cite{moelmer,amid}, being nonzero and even maximal for states approaching the classical limit, and thus severely overestimating quantum correlations. To date, the MID has not been computed for Gaussian states.

{\it Classical mutual information and ameliorated measurement-induced disturbance.}--- To cure this major drawback of MID, one can define an `ameliorated measurement-induced disturbance' (AMID) by incorporating into \eq{MID2} an optimization (precisely, a minimization) over the joint bi-local POVM measurement $\hat{\Pi}_{A}\otimes\hat{\Pi}_{B}$ on subsystems $A$ and $B$ \cite{amid}. The AMID then can be defined as \cite{piani,moelmer,amid}
\begin{eqnarray}\label{AMID}
{\cal A}(\hat{\rho}_{AB})&=&\inf_{\hat{\Pi}_{A}\otimes\hat{\Pi}_{B}}\left\{{\cal I}_q(\hat{\rho}_{AB})-{\cal I}(A:B)\right\} \\
&=&{\cal I}_{q}(\hat{\rho}_{AB})-{\cal I}_{c}(\hat{\rho}_{AB})\,, \nonumber
\end{eqnarray}
where
%%%%%%%%%%%%%%%%%%%%%%%%%%%%%%%%%%%%%%%%%%%%%%%%%%%%%%%%%%%%%%%%%%%%%%%%%%%%%%%%%%%%%%%%%%%%%%%%%%%
\begin{eqnarray}\label{Ic}
{\cal I}_{c}(\hat{\rho}_{AB})=\sup_{\hat{\Pi}_{A}\otimes\hat{\Pi}_{B}}{\cal I}(A:B)
\end{eqnarray}
%%%%%%%%%%%%%%%%%%%%%%%%%%%%%%%%%%%%%%%%%%%%%%%%%%%%%%%%%%%%%%%%%%%%%%%%%%%%%%%%%%%%%%%%%%%%%%%%%%%
is the classical mutual information of a quantum state $\hat{\rho}_{AB}$ \cite{terhal}, where ${\cal I}(A:B)$
is the classical mutual information of the joint probability distribution
$p_{AB}(k,l)=\mbox{Tr}[\hat{\rho}_{AB}\hat{\Pi}_{A}(k)\otimes\hat{\Pi}_{B}(l)]$ of outcomes of local measurements
$\hat{\Pi}_{A}$ and $\hat{\Pi}_{B}$ on $\hat{\rho}_{AB}$.
The AMID captures the quantumness of bipartite correlations as signaled by the minimal state disturbance after optimized local measurements. As such, it is a symmetric, strongly faithful nonclassicality measure \cite{amid} that vanishes if and only if a bipartite state $\hat{\rho}_{AB}$ is genuinely classically correlated \cite{piani,moelmer}, and it is operationally interpreted as the quantum complement to the classical mutual information [\eq{Ic}], while the latter is in turn a {\it bona fide} measure of classical correlations in general bipartite quantum states \cite{terhal}. The AMID thus englobes the nice properties of discord and MID without showing their genetic weaknesses \cite{amid,brasilesymm}.
 The evaluation and properties of AMID have been recently investigated for two--qubit systems \cite{amid}.  \footnote{In Ref.~\cite{amid} the AMID is defined via an optimization over local projective measurements, rather than more general local POVMs. Both versions of AMID have also been studied in Ref.~\cite{moelmer}, albeit without naming the considered measures explicitly.}

{\it Hierarchy of nonclassical correlations.}---The three entropic nonclassicality indicators introduced above (discord, MID and AMID) satisfy the following hierarchical relationship on arbitrary bipartite quantum states \cite{moelmer,amid}: \begin{equation}
\label{hierardy}
\{{\cal D}^{\leftarrow},{\cal D}^{\rightarrow}\} \le {\cal D}^{\leftrightarrow} \le {\cal A} \le {\cal M}\,.
 \end{equation}

{\it Entanglement of Formation.}---
 For pure bipartite states $\ket{\psi}_{AB}$,  all the measures of nonclassical correlations introduced above (discord, MID and AMID) reduce to the canonical entanglement measure,  the `entropy of entanglement', \begin{equation}\label{EE}
E(\ket{\psi}_{AB}) = {\cal S}(\mbox{Tr}_B \ket{\psi}\!_{AB}\bra{\psi})\,.
 \end{equation}
This shows that quantum correlations are faithfully identified with just entanglement in the special case of pure states of composite quantum systems.

For bipartite mixed states  $\hat{\rho}_{AB}$, let us recall that the Entanglement of Formation (EoF) $E_f(\hat{\rho}_{AB})$ is a well-known entanglement monotone \cite{bennet96}, defined as the convex roof of the pure-state entropy of entanglement
\begin{equation}\label{EOF}
E_{f}(\hat{\rho}_{AB})=\min_{\{p_i,\ket{\psi}_{AB}^i\}}\sum_i p_i E(\ket{\psi}_{AB}^i)
\;, \end{equation}
where the minimum is taken over all the pure-state realizations of $\hat{\rho}_{AB}$,
\[
\hat{\rho}_{AB}=\sum_i p_i \ket{\psi}_{AB}^i \bra{\psi}_{AB}^i \; .
\]

For general mixed bipartite Gaussian states $\hat{\rho}_{AB}$, one can introduce a Gaussian version of the EoF (Gaussian EoF) $E_f^G$, which is by construction an upper bound to $E_f$, defined as the convex roof of the entropy of entanglement restricted to decompositions of $\hat{\rho}_{AB}$ into pure Gaussian states only \cite{geof}. The Gaussian EOF can be evaluated via a minimization over CMs: \begin{equation}\label{geof}
{E}_f^G(\gamma) = \inf_{\gamma' \le \gamma\ :\ {\det(\gamma')=1}} E(\gamma')\,,\end{equation}
where the infimum runs over all pure bipartite Gaussian states with CM $\gamma'$ smaller than $\gamma$. Compact formulas for $E_f^G$ exist for all symmetric two--mode states  (where $E_f^G=E_f$ as the Gaussian decomposition is globally optimal) \cite{giedke03}, and in the nonsymmetric case for squeezed thermal states and so-called states of partial minimum uncertainty \cite{ordering}; for all the other two--mode states its value can still be found analytically \cite{geof}.

In the following Sections, we will approach the evaluation and characterization of MID and AMID for arbitrary two--mode Gaussian states and their interplay with entanglement, complementing the studies of Refs.~\cite{Giorda_10,AD_10} on discord.

\section{Measurement-induced disturbance of two--mode Gaussian states}\label{secMID}

In this Section we will address the calculation of the  MID ${\cal M}$ \cite{MID}, \eq{MID}, for two--mode Gaussian states. Here and in the following we will always consider, without any loss of generality, states $\hat{\rho}_{AB}$ whose CM $\gamma$ is in standard form, \eq{gamma}. The expressions we derive can be straightforwardly recast in terms of a set of four local symplectic invariants for general two--mode states, that uniquely define the standard form covariances \cite{Simon_00,extremal}.

The quantum mutual information of a two--mode Gaussian state can be computed via the formula (\ref{MIG}). Here we will be concerned with the calculation of the mutual information after local projections onto the eigenstates of the marginal density matrices.

%%%%%%%%%%%%%%%%%%%%%%%%%%%%%%%%%%%%%%%%%%%%%%%%%%%%%%%%%%%%%%%%%%%%%%%%%%%%%%%%%%
For a generic two--mode mixed Gaussian state
$\hat{\rho}_{AB}$ in standard form, the reduced states
are just thermal states $\hat{\rho}_{{\rm th},A}$ and $\hat{\rho}_{{\rm th},B}$ with mean
number of thermal photons $\langle \hat{n}_{A}\rangle=(a-1)/2$ and $\langle \hat{n}_{B}\rangle=(b-1)/2$,
respectively. The local measurements $\hat{\cal E}_A \otimes \hat{\cal E}_B$ entering the expression of MID,  \eq{MID},
are then {\it non-Gaussian} measurements, given by projections onto Fock states (joint photon counting),
%%%%%%%%%%%%%%%%%%%%%%%%%%%%%%%%%%%%%%%%%%%%%%%%%%%%%%%%%%%%%%%%%%%%%%%%%%%%%%%%%%%%
\begin{eqnarray}\label{Fock}
\hat{\cal E}_{j}(n)=|n\rangle_{j}\langle n|,\quad j=A,B,
\end{eqnarray}
%%%%%%%%%%%%%%%%%%%%%%%%%%%%%%%%%%%%%%%%%%%%%%%%%%%%%%%%%%%%%%%%%%%%%%%%%%%%%%%%%%%%
and the post-measurement state reads as
%%%%%%%%%%%%%%%%%%%%%%%%%%%%%%%%%%%%%%%%%%%%%%%%%%%%%%%%%%%%%%%%%%%%%%%%%
\begin{equation}\label{Pirho}
\hat{\cal E}(\hat{\rho}_{AB})=\sum_{m,n=0}^{\infty}p(m,n)|m\rangle_{A}\langle m|\otimes|n\rangle_{B}\langle n|,
\end{equation}
%%%%%%%%%%%%%%%%%%%%%%%%%%%%%%%%%%%%%%%%%%%%%%%%%%%%%%%%%%%%%%%%%%%%%%%%%
where $p(m,n)=\!\!_A\langle m|_{B}\langle n|\hat{\rho}_{AB}|m\rangle_{A}|n\rangle_{B}$ is the joint probability
distribution of finding $m$ photons in mode $A$ and $n$ photons in mode $B$. %(subscript $AB$ is omitted for simplicity in what follows).
The determination of the quantum mutual information ${\cal I}_{q}[\hat{\cal E}(\hat{\rho}_{AB})]$ of the
post-measurement state (\ref{Pirho}) requires the calculation of the local von Neumann
entropies ${\cal S}(\hat{\rho}_{j}^{\cal E})$, $j=A,B$ of the reduced states
$\hat{\rho}_{A,B}^{\cal E}=\mbox{Tr}_{B,A}\left[\hat{\cal E}(\hat{\rho}_{AB})\right]$ as well as
the global entropy ${\cal S}[\hat{\cal E}(\hat{\rho}_{AB})]$. Since the reduced states are just
equal to the local thermal states, i.e. $\hat{\rho}_{j}^{\cal E}=\hat{\rho}_{{\rm th},j}$, $j=A,B$,
we obtain immediately the local entropies equal to ${\cal S}(\hat{\rho}_{A}^{\cal E})=\mathfrak{F}(a)$
and ${\cal S}(\hat{\rho}_{B}^{\cal E})=\mathfrak{F}(b)$, where $\mathfrak{F}(x)$ is defined by \eq{effe}.
The global entropy can be computed from the joint photon-number probability distribution $p(m,n)$. The latter
can be derived using the generating function for the distribution in the spirit of Ref.~\cite{Perina_05}.

We start by noting that any two--mode quantum state $\hat{\rho}_{AB}$ can be
represented by the complex normal quantum characteristic function defined as
%%%%%%%%%%%%%%%%%%%%%%%%%%%%%%%%%%%%%%%%%%%%%%%%%%%%%%%%%%%%%%%%%%%%%%%%%%%%%%%%%%%%%%%%%%%%%%%%
\begin{eqnarray}\label{CNdefinition}
C(\beta_{1},\beta_{2})=\mbox{Tr}[\hat{\rho}_{AB}e^{\beta_{1}\hat{a}^{\dag}+\beta_{2}\hat{b}^{\dag}}
e^{-\beta_{1}^{\ast}\hat{a}-\beta_{2}^{\ast}\hat{b}}],
\end{eqnarray}
%%%%%%%%%%%%%%%%%%%%%%%%%%%%%%%%%%%%%%%%%%%%%%%%%%%%%%%%%%%%%%%%%%%%%%%%%%%%%%%%%%%%%%%%%%%%%%%%
where $\hat{a}$ $(\hat{a}^{\dag})$ and $\hat{b}$ $(\hat{b}^{\dag})$ are annihilation (creation)
operators of modes $A$ and $B$, and $\beta_{1},\beta_{2}$ are complex parameters of the characteristic function.
For a Gaussian state in the standard form (\ref{gamma}) with zero means $\langle\hat{a}\rangle=\langle\hat{b}\rangle=0$
the characteristic function attains the form
%%%%%%%%%%%%%%%%%%%%%%%%%%%%%%%%%%%%%%%%%%%%%%%%%%%%%%%%%%%%%%%%%%%%%%%%%%%%%%%%%%%%%%%%%%%%%%%%
\begin{eqnarray}\label{CN}
C(\beta_{1},\beta_{2})&=&\exp\left[-(B_{1}|\beta_{1}|^2+B_{2}|\beta_{2}|^2)+(D\beta_{1}^{\ast}\beta_{2}^{\ast}\right.\nonumber\\
&&\left.+\bar{D}\beta_{1}\beta_{2}^{\ast}+\mbox{c.c.})\right],
\end{eqnarray}
%%%%%%%%%%%%%%%%%%%%%%%%%%%%%%%%%%%%%%%%%%%%%%%%%%%%%%%%%%%%%%%%%%%%%%%%%%%%%%%%%%%%%%%%%%%%%%%%
where $B_{1}=\langle\Delta \hat{a}^{\dag}\Delta\hat{a}\rangle=(a-1)/2$, $B_{2}=\langle\Delta \hat{b}^{\dag}\Delta\hat{b}\rangle=(b-1)/2$,
$D=\langle\Delta \hat{a}\Delta\hat{b}\rangle=(c_{1}-c_{2})/4$, $\bar{D}=-\langle\Delta \hat{a}^{\dag}\Delta\hat{b}\rangle=-(c_{1}+c_{2})/4$.
Here c.c. stands for complex conjugate terms and $\Delta\hat{A}\equiv\hat{A}-\langle\hat{A}\rangle$. The generating function for the distribution $p(m,n)$ then reads
\cite{Perina_91}
%%%%%%%%%%%%%%%%%%%%%%%%%%%%%%%%%%%%%%%%%%%%%%%%%%%%%%%%%%%%%%%%%%%%%%%%%%%%%%%%%%%%%%%%%%%%%%%%%
\begin{eqnarray}\label{G}
G\left(\lambda_{1},\lambda_{2}\right)&=&\frac{1}{\pi^2\lambda_{1}\lambda_{2}}\int\int\exp\left(-\frac{|\beta_{1}|^{2}}{\lambda_{1}}-
\frac{|\beta_{2}|^{2}}{\lambda_{2}}\right)\nonumber\\
&&C(\beta_{1},\beta_{2}){\rm d}^{2}\beta_{1}{\rm d}^{2}\beta_{2},
\end{eqnarray}
%%%%%%%%%%%%%%%%%%%%%%%%%%%%%%%%%%%%%%%%%%%%%%%%%%%%%%%%%%%%%%%%%%%%%%%%%%%%%%%%%%%%%%%%%%%%%%%%%
where $\lambda_{1}$ and $\lambda_{2}$ are real parameters. By inserting the characteristic function (\ref{CN}) into
the integral (\ref{G}) and performing the integration, we arrive at the generating function
$G\left(\lambda_{1},\lambda_{2}\right)=F_{1}\left(\lambda_{1},\lambda_{2}\right)F_{2}\left(\lambda_{1},\lambda_{2}\right)$,
with
%%%%%%%%%%%%%%%%%%%%%%%%%%%%%%%%%%%%%%%%%%%%%%%%%%%%%%%%%%%%%%%%%%%%%%%%%%%%%%%%%%%%%%%%%%%%%%%%
\begin{eqnarray}\label{Gj}
F_{j}\left(\lambda_{1},\lambda_{2}\right)=\frac{1}{\sqrt{1+B_{1}\lambda_{1}+B_{2}\lambda_{2}+K_{j}\lambda_{1}\lambda_{2}}},
\end{eqnarray}
%%%%%%%%%%%%%%%%%%%%%%%%%%%%%%%%%%%%%%%%%%%%%%%%%%%%%%%%%%%%%%%%%%%%%%%%%%%%%%%%%%%%%%%%%%%%%%%%
where $K_{j}=[(a-1)(b-1)-c_{j}^2]/4$, $j=1,2$. The sought photon-number distribution is obtained by differentiating
the generating function (\ref{G}) as
%%%%%%%%%%%%%%%%%%%%%%%%%%%%%%%%%%%%%%%%%%%%%%%%%%%%%%%%%%%%%%%%%%%%%%%%%%%%%%%%%%%%%%%%%%%%%%%%%%
\begin{eqnarray}\label{pmnfromG}
p(m,n)&=&{\frac{(-1)^{m+n}}{m!n!}\frac{\partial^{m+n}G\left(\lambda_{1},\lambda_{2}\right)}
{\partial\lambda_{1}^{m}\partial\lambda_{2}^{n}}\vline}_{\lambda_{1}=\lambda_{2}=1},
\end{eqnarray}
%%%%%%%%%%%%%%%%%%%%%%%%%%%%%%%%%%%%%%%%%%%%%%%%%%%%%%%%%%%%%%%%%%%%%%%%%%%%%%%%%%%%%%%%%%%%%%%%%%
which gives us explicitly
%%%%%%%%%%%%%%%%%%%%%%%%%%%%%%%%%%%%%%%%%%%%%%%%%%%%%%%%%%%%%%%%%%%%%%%%%%%%%%%%%%%%%%%%%%%%%%%%%%
\begin{eqnarray}\label{pmn}
p(m,n)&=&\frac{1}{m!n!}\sum_{\nu_{1}=0}^{m}\sum_{\nu_{2}=0}^{n}{m \choose \nu_{1}}{n \choose \nu_{2}}Q^{(1)}(\nu_{1},\nu_{2})\nonumber\\
&&\times Q^{(2)}(m-\nu_{1},n-\nu_{2}),
\end{eqnarray}
%%%%%%%%%%%%%%%%%%%%%%%%%%%%%%%%%%%%%%%%%%%%%%%%%%%%%%%%%%%%%%%%%%%%%%%%%%%%%%%%%%%%%%%%%%%%%%%%%%
where
%%%%%%%%%%%%%%%%%%%%%%%%%%%%%%%%%%%%%%%%%%%%%%%%%%%%%%%%%%%%%%%%%%%%%%%%%%%%%%%%%%%%%%%%%%%%%%%%%%
\begin{eqnarray}\label{Q}
Q^{(j)}(\alpha,\beta)&=&\frac{(B_{1}+K_{j})^{\alpha}(B_{2}+K_{j})^{\beta}}{4^{\alpha+\beta}(1+B_{1}+B_{2}+K_{j})^{\alpha+\beta+\frac{1}{2}}}\nonumber\\
&&\times\sum_{l=0}^{\mbox{\small min}(\alpha,\beta)}l!{\alpha \choose l}{\beta \choose l}\frac{[2(\alpha+\beta-l)]!}{(\alpha+\beta-l)!}\nonumber\\
&&\times\left[-4K_{j}\frac{1+B_{1}+B_{2}+K_{j}}{(B_{1}+K_{j})(B_{2}+K_{j})}\right]^{l}.
\end{eqnarray}
%%%%%%%%%%%%%%%%%%%%%%%%%%%%%%%%%%%%%%%%%%%%%%%%%%%%%%%%%%%%%%%%%%%%%%%%%%%%%%%%%%%%%%%%%%%%%%%%%%%
The global posterior von Neumann entropy is then given by the formula \begin{equation}
\label{}
{\cal S}[\hat{\cal E}(\hat{\rho}_{AB})]=-\sum_{m,n=0}^{\infty}p(m,n)\ln p(m,n)
\end{equation}
and can be evaluated numerically for arbitrary two--mode Gaussian states.

We note that in Refs.~\cite{Dodonov} it is shown that the joint photon-number distribution for
multimode Gaussian states can be written in terms of multivariable Hermite polynomials.
Recursive formulas can be derived for the calculation of the multivariable Hermite Polynomials,
and this fact can be exploited to speed up the numerical evaluation of the MID.

The final expression for the MID of two--mode Gaussian states $\hat\rho_{AB}$ with a standard form CM as in \eq{gamma} is then
\begin{eqnarray}\label{MIDG}
{\cal M}(\hat\rho_{AB}) &=& {\cal S}[\hat{\cal E}(\hat{\rho}_{AB})] - {\cal S}(\hat{\rho}_{AB}) \\
&=& -\sum_{m,n=0}^{\infty}p(m,n)\ln p(m,n) - \mathfrak{F}(\nu_+)-\mathfrak{F}(\nu_-) \nonumber \,.
\end{eqnarray}

%A comparison between MID ${\cal M}$, defined by \eq{MIDG}, and two-way Gaussian discord ${\cal D}^{\leftrightarrow}$, \eq{twoway}, evaluated in Ref.~\cite{AD_10}, is presented for random two--mode Gaussian states in Fig.~\ref{azz2}. We notice a behavior very similar to the corresponding case for two--qubit states (cfr. Fig. 2(a) in Ref.~\cite{amid}): MID is a very loose upper bound to quantum discord, equating it on pure states, and drastically overestimating quantum correlations on general mixed states, with the possibility of reaching arbitrarily large values even on states with infinitesimal discord. This feature will be further commented on in Section \ref{secDiscuss}.

In the subcase of two--mode squeezed  states, with $c_{1}=\pm c_{2}=c$,
we have $K_{1}=K_{2}=K=[(a-1)(b-1)-c^2]/4$ and the distribution (\ref{pmn}) takes the simplified form
%%%%%%%%%%%%%%%%%%%%%%%%%%%%%%%%%%%%%%%%%%%%%%%%%%%%%%%%%%%%%%%%%%%%%%%%%%%%%%%%%%%%%%%%%%%%%%%%%%
\begin{eqnarray}\label{pmnsqueezedthermal}
p(m,n)&=&\frac{(B_{1}+K)^{m}(B_{2}+K)^{n}}{m!n!(1+B_{1}+B_{2}+K)^{m+n+1}}\nonumber\\
&&\times\sum_{j=0}^{\mbox{\small min}(m,n)}{m \choose j}{n \choose j}j!(m+n-j)!\nonumber\\
&&\times\left[-K\frac{1+B_{1}+B_{2}+K}{(B_{1}+K)(B_{2}+K)}\right]^{j}.
\end{eqnarray}
%%%%%%%%%%%%%%%%%%%%%%%%%%%%%%%%%%%%%%%%%%%%%%%%%%%%%%%%%%%%%%%%%%%%%%%%%%%%%%%%%%%%%%%%%%%%%%%%%%

For the very special instance of pure two--mode Gaussian states with the standard form given by
two--mode squeezed vacuum states
 \begin{equation}\label{TMSV}
|\psi(r)\rangle_{AB}=\sqrt{1-q^2}\sum_{n=0}^{\infty}q^{n}|n,n\rangle_{AB},
\end{equation}
 with standard form covariances given by \eq{puri}, where $q=\tanh r$,  the MID can be evaluated in closed form.
  The state is in Schmidt decomposition
with Schmidt coefficients $\lambda_{n}=\sqrt{1-q^2}q^{n}$. Being a pure state, ${\cal S}(|\psi(r)\rangle_{AB})=0$.
The post-measurement state (\ref{Pirho}) then reads \cite{MID}
%%%%%%%%%%%%%%%%%%%%%%%%%%%%%%%%%%%%%%%%%%%%%%%%%%%%%%%%%%%%%%%%%%%%%%%%%
\begin{equation}\label{Pirhopuri}
\hat{\cal E}(|\psi(r)\rangle_{AB})=(1-q^{2})\sum_{n=0}^{\infty}q^{2n}|n\rangle_{A}\langle n|\otimes|n\rangle_{B}\langle n|,
\end{equation}
%%%%%%%%%%%%%%%%%%%%%%%%%%%%%%%%%%%%%%%%%%%%%%%%%%%%%%%%%%%%%%%%%%%%%%%%%
and its  von Neumann entropy is  ${\cal S}[\hat{\cal E}(|\psi(r)\rangle_{AB})]=-\sum_{n}\lambda_{n}^2\ln\lambda_n^2 = \mathfrak{F}[\cosh(2r)]$, which precisely coincides with the entropy of entanglement of the pure two--mode Gaussian state, \eq{EE}. Therefore,
\begin{eqnarray}\label{MIDpuri}
{\cal M}[|\psi(r)\rangle_{AB}]&=& E[|\psi(r)\rangle_{AB}]=\cosh^{2}(r)\ln[\cosh^{2}(r)]\nonumber\\
&&-\sinh^{2}(r)\ln[\sinh^{2}(r)]\,,
\end{eqnarray}
as expected from the definition of MID. Let us remark that this value is attained by local non-Gaussian measurements (joint photon counting). An interesting question is whether there exist local Gaussian POVMs that can result in  a measurement-induced disturbance equal to the one in \eq{MIDpuri} for pure two--mode Gaussian states. Surprisingly enough, we will prove in the next Section that the answer is negative.

\section{Gaussian ameliorated measurement-induced disturbance of two--mode Gaussian states} \label{secGAMID}

Given the crucial role of Gaussian states in quantum information processing, it is important to carry out a thorough analysis and comprehensive characterization of their quantum and classical correlations. A primal step to undertake is to approach the evaluation of faithful measures, such as the classical mutual information ${\cal I}_c$ and correspondingly the AMID ${\cal A}$ [Eqs.~(\ref{AMID},\ref{Ic})],  for arbitrary two--mode Gaussian states $\hat{\rho}_{AB}$. However, the problem appears formidable if any possible non-Gaussian POVM is allowed. Therefore, similarly to what has been done for discord \cite{Giorda_10,AD_10}, one can define a Gaussian version of the AMID, ${\cal A}^G$, where the optimization is constrained to local Gaussian POVMs of the form (\ref{POVM}), as
\begin{equation}\label{GAMID}
{\cal A}^{G}(\hat{\rho}_{AB})={\cal I}_q(\hat{\rho}_{AB})-{\cal I}_{c}^{G}(\hat{\rho}_{AB}),
\end{equation}
where
\begin{equation}\label{IcG}
{\cal I}_{c}^{G}(\hat{\rho}_{AB})=\sup_{\hat{\Pi}_{A}^G\otimes\hat{\Pi}_{B}^G}{\cal I}(A:B)\,
\end{equation}
is the Gaussian classical mutual information of the quantum state $\hat{\rho}_{AB}$.
%\begin{equation}\label{GAMID}
%{\cal A}^{G}(\hat{\rho}_{AB})={\cal I}_q(\hat{\rho}_{AB})-\underbrace{\sup_{\hat{\Pi}_{A}^G\otimes\hat{\Pi}_{B}^G}{\cal I}(A:B)}_{\text{Gaussian\ } %{\cal I}_c}\,.
%\end{equation}
The true AMID, optimized over general local measurements, would be then bounded from above as \begin{equation}\label{boundamid}
{\cal A}(\hat\rho_{AB}) \le \min\{{\cal A}^G(\hat\rho_{AB}), {\cal M}(\hat\rho_{AB})\}.
\end{equation}

A relevant and natural question, related to the one in the closing of the previous Section, is whether Gaussian measurements are always optimal for the evaluation of the AMID, i.e., whether ${\cal A} ={\cal A}^G$ for all bipartite Gaussian states. In the case of discord, there are subclasses of two--mode Gaussian states where one can prove that the Gaussian discord achieves the global minimum in \eq{discord} even when including potentially non-Gaussian measurements on a local subsystem \cite{AD_10}: the properties of the Gaussian discord allow us  to conjecture that this might be true for arbitrary two--mode Gaussian states, although no rigorous investigation is available to date to support this claim \footnote{A preliminary numerical study reveals that, for general two--mode mixed Gaussian states, even allowing for a non-Gaussian measurement such as photon counting, the corresponding discord is found to be never smaller than the one associated to the optimal Gaussian measurement [A. Datta, private communication].}. Remarkably, we will find instead that when both parties are locally probed,  there is a finite-volume set of two--mode Gaussian states (notably including pure states) for which non-Gaussian measurements are necessarily optimal for the evaluation of the AMID, and the Gaussian AMID only provides a strict upper bound to it. This proves that Gaussian joint measurements may not be the least disturbing ones on general two--mode Gaussian states.

We hereby develop the framework for the determination of the Gaussian AMID on general two--mode Gaussian states, and provide closed formulas for it in some special cases.
The nontrivial part in the determination of the quantity (\ref{GAMID}) is the calculation of the classical
mutual information ${\cal I}_{c}^G(\hat{\rho}_{AB})$ requiring maximization of the Shannon mutual information
${\cal I}(A:B)$ over local Gaussian POVMs $\hat{\Pi}_{A}$ and $\hat{\Pi}_{B}$ of the form (\ref{POVM}).

We begin by observing that in the considered optimization task we can restrict ourselves to covariant Gaussian POVMs (\ref{POVM}) projecting onto pure states (rank-one POVMs), similarly to discrete-variable scenarios \cite{moelmer}. The proof of this statement is provided in Appendix \ref{secAppPOVM}.
We can thus focus on local measurements of the form (\ref{POVM}), where the seed state $\hat\Pi_j$ is a pure single--mode Gaussian state with CM $\gamma_j$. Let us recall that any pure-state one--mode
CM $\gamma_{j}$ can be expressed as $\gamma_{j}=U(\theta_{j})V(r_{j})U^{T}(\theta_{j})$, where
%%%%%%%%%%%%%%%%%%%%%%%%%%%%%%%%%%%%%%%%%%%%%%%%%%%%%%%%%%%%%%%%%%%%%%%%%%%%%%%%%%%%%%%%%%%%%%%%%%%%%%%%%%%
\begin{equation}\label{UV}
U(\theta_{j})= \left(
\begin{array}{cc}
\cos\theta_{j}  & \sin \theta_{j} \\
-\sin\theta_{j} & \cos\theta_{j}
\end{array}
\right),
\quad
V(r_{j})= \left(
\begin{array}{cc}
e^{2r_{j}}  & 0 \\
0 & e^{-2r_{j}}
\end{array}
\right),
\end{equation}
%%%%%%%%%%%%%%%%%%%%%%%%%%%%%%%%%%%%%%%%%%%%%%%%%%%%%%%%%%%%%%%%%%%%%%%%%%%%%%%%%%%%%%%%%%%%%%%%%%%%%%%%%%%%%
where $\theta_{j}\in\langle0,\pi)$ and $r_{j}\geq0$. In this picture, {\it homodyne detection} on mode $j$ is recovered
in the limit of an infinitely squeezed pure state $\hat{\Pi}_j$, i.e.,~$r_j \rightarrow \infty$.
On the other hand, {\it heterodyne detection} on mode $j$ corresponds to $r_j=0$.

Now we want to maximize the Shannon mutual information ${\cal I}(A:B)$ of the distribution (\ref{Pd}),
$$P(d)=\mbox{Tr}[\hat{\Pi}_{A}(d_{A})\otimes\hat{\Pi}_{B}(d_{B})\hat{\rho}_{AB}],$$
over all single--mode pure-state CMs $\gamma_{A,B}$.
Expressing the two--mode state CM $\gamma$ in block form as in \eq{gamma},
and using the formula for the Shannon entropy of a Gaussian distribution $P$ of $N$ variables with classical correlation matrix $\Sigma$,
${\cal H}(P)=\ln[(2\pi e)^{\frac{N}{2}}\sqrt{\det\Sigma}]$ \cite{Shannon_48}, the sought mutual information
can be obtained in the form \cite{Gelfand_57}:
%%%%%%%%%%%%%%%%%%%%%%%%%%%%%%%%%%%%%%%%%%%%%%%%%%%%%%%%%%%%%%%%%%%%%%%%%%%%%%%%%%%%%%%%%%%%%%%%%%%%%%%%
\begin{equation}\label{IAB}
{\cal I}(A:B)=\frac{1}{2}\ln\left[\frac{\det\left(\gamma_{A}+A\right)\det\left(\gamma_{B}+B\right)}
{\det\left(\gamma_{A}\oplus\gamma_{B}+\gamma\right)}\right].
\end{equation}
%%%%%%%%%%%%%%%%%%%%%%%%%%%%%%%%%%%%%%%%%%%%%%%%%%%%%%%%%%%%%%%%%%%%%%%%%%%%%%%%%%%%%%%%%%%%%%%%%%%%%%%%%

Since the determinant is invariant with respect to symplectic transformations we can assume the CM $\gamma$ to be in
standard form, \eq{gamma}; moreover, given \eq{effe}, \eq{UV}, the invariance of the determinant under orthogonal transformations, and the monotonicity of the logarithmic function, the object to be maximized
reads
%%%%%%%%%%%%%%%%%%%%%%%%%%%%%%%%%%%%%%%%%%%%%%%%%%%%%%%%%%%%%%%%%%%%%%%%%%%%%%%%%%%%%%%%%%%%%%%%%%%%%%%%%%%%%
\begin{equation}\label{f}
f(r_{A},r_{B},\theta_{A},\theta_{B})=\frac{\det{A'}\det{B'}}
{\det\gamma'},
\end{equation}
%%%%%%%%%%%%%%%%%%%%%%%%%%%%%%%%%%%%%%%%%%%%%%%%%%%%%%%%%%%%%%%%%%%%%%%%%%%%%%%%%%%%%%%%%%%%%%%%%%%%%%%%%%%%%%
where the CM $\gamma'$ has the $2\times2$ blocks of the form
\begin{equation}\label{blockprime}
\begin{split}
&A'=a\openone+V(r_{A})\,,\quad B'=b\openone+V(r_{B})\,, \\
&C'=U^{T}(\theta_{A})\mbox{diag}(c_{1},c_{2})U(\theta_{B})\,.
 \end{split}\end{equation}
We recall that the Gaussian AMID takes then the form (\ref{GAMID}) with
\begin{equation}\label{GAMIDG}
{\cal I}_{c}^{G}(\hat{\rho}_{AB})=\frac12 \ln \left[\sup_{\{r_{A,B},\theta_{A,B}\}} f(r_{A},r_{B},\theta_{A},\theta_{B})\right]\,.
\end{equation}
 The determinant in the denominator of \eq{f} can be expressed in terms of the local symplectic invariants $I_{1}=\det A'$, $I_{2}=\det B'$,
$I_{3}=\det C'$ and $I_{4}=\mbox{Tr}(A'JC'JB'JC'^{T}J)$ as $\det\gamma'=I_1I_2+I_3^{2}-I_{4}$ \cite{Simon_00}.
Evidently, the function (\ref{f}) depends on the phases $\theta_{A,B}$ only through invariant $I_{4}$ so we can optimize
$f$ over the phases by optimizing $I_{4}$. Since $I_{4}$ is always nonnegative, the maximum of the function $f$ will
be obtained if the phases will maximize $I_{4}$.

Three different cases must be distinguished depending on the values of the
eigenvalues $c_{1,2}$ of the matrix $C$.\\

\begin{enumerate}
  \item If $c_{1}=c_{2}=0$, then $I_{3,4}=0$, which implies $f=1$ and therefore ${\cal I}_{c}^{G}(\hat{\rho}_{AB})=0$.
Since also ${\cal I}_{q}(\hat{\rho}_{AB})=0$, we get finally ${\cal A}^G(\hat{\rho}_{AB})=0$. Obviously, product states have neither quantum nor classical correlations. Actually, it is known that any non-product two--mode Gaussian state has nonzero quantum correlations, so we must expect be ${\cal A}^G > 0$ as soon as one of the covariances in the block $C$ is nonzero.

  \item If $c_{1}>0$ and $c_{2}=0$,  then for $r_{A,B}>0$ the optimal phases are $\theta_{A,B,{\rm opt}}=\pi/2$ and for these phases the invariant $I_4$ takes the value $I'_{4}=c_{1}^{2}(a+e^{2r_{A}})(b+e^{2r_{B}})$. Let us define
      \begin{eqnarray}\label{gviah}
      \frac{1}{1-h(r_{A},r_{B})}&\equiv& g(r_{A},r_{B})\nonumber\\
      &\equiv& f(r_{A},r_{B},\theta_{A,{\rm opt}},\theta_{B,{\rm opt}}).
      \end{eqnarray}
We have $h=I_{4}'/(I_{1}I_{2})$ and the function $h$ (and consequently $g$) is obviously maximized
in the limits $r_{A,B}\rightarrow\infty$ (doubly homodyne detection) when we get
%%%%%%%%%%%%%%%%%%%%%%%%%%%%%%%%%%%%%%%%%%%%%%%%%%%%%%%%%%%%%%%%%%%%%%%%%%%%%%%%%%%%%%%%%%%%%%%%%%%%%%%%%%%%%%%%%
\begin{equation}\label{ghom}
g_{\rm hom}=\frac{ab}{ab-c_{1}^{2}}.
\end{equation}
%%%%%%%%%%%%%%%%%%%%%%%%%%%%%%%%%%%%%%%%%%%%%%%%%%%%%%%%%%%%%%%%%%%%%%%%%%%%%%%%%%%%%%%%%%%%%%%%%%%%%%%%%%%%%%%%%
Since $g_{\rm hom}$ is also larger than the maximal values of the function $g$ on the boundaries
$r_{A}=0$ or $r_{B}=0$, we arrive at the conclusion that (\ref{ghom}) is the optimal value of \eq{f}
for states with $c_{2}=0$ \footnote{Recall that in our convention $c_1 \ge |c_2|$, therefore the optimal
Gaussian measurement consists in the local homodyne detection of the quadrature with the highest intermodal correlation.}.
%%%%%%%%%%%%%%%%%%%%%%%%%%%%%%%%%%%%%%%%%%%%%%%%%%%%%%%%%%%%%%%%%%%%%%%%%%%%%%%%%%%%%%%%%
%%%%%%%%%%%%%%%%%%%%%%%%%%%%%%%%%%%%%%%%%%%%%%%%%%%%%%%%%%%%%%%%%%%%%%%%%%%%%%%%%%%%%%%%%

  \item
  In the general case $c_{1},c_{2}\ne 0$, the invariant $I_{4}$ can be expressed as $I_{4}=c_{1}^{2}c_{2}^{2}\mbox{Tr}(XB')$,
where $X$ is a real symmetric positive-semidefinite matrix of the form:
$X=U_{B}^{T}\mbox{diag}(c_{1}^{-1},c_{2}^{-1})U_{A}A'U_{A}^{T}\mbox{diag}(c_{1}^{-1},c_{2}^{-1})U_{B}$, with
$U_{j}\equiv U(\theta_{j})$  defined in Eq.~(\ref{UV}), and the diagonal matrices $A'$ and $B'$ defined in \eq{blockprime}.
Expressing the matrix $X$ through eigenvalue decomposition $X=W(\phi)\mbox{diag}[x_{1}(\theta_{A}),x_{2}(\theta_{A})]W^{T}(\phi)$,
where $W$ is an orthogonal matrix diagonalizing $X$ and $x_{1}(\theta_{A})\geq x_{2}(\theta_{A})$ are the eigenvalues of $X$ depending on the angle $\theta_{A}$, we can  maximize $I_{4}$ over the phase $\phi$. Further, if we substitute into the obtained formula the explicit forms of eigenvalues $x_{1,2}(\theta_{A})$, we can perform also the maximization over the angle $\theta_{A}$, which finally yields the invariant
$I_{4}$ maximized over the phases $\theta_{A,B}$ of local measurements of the form:
%%%%%%%%%%%%%%%%%%%%%%%%%%%%%%%%%%%%%%%%%%%%%%%%%%%%%%%%%%%%%%%%%%%%%%%%%%%%%%%%%%%%%%%%%%%%%%%%
\begin{eqnarray}\label{I4primed}
I'_{4}&=&\left\{\left[a+\cosh\left(2r_{A}\right)\right]\left[b+\cosh\left(2r_{B}\right)\right]\right.\nonumber\\
&&\left.+\sinh\left(2r_{A}\right)\sinh\left(2r_{B}\right)\right\}\left(c_{1}^{2}+c_{2}^{2}\right)\nonumber\\
&&+\left\{\left[a+\cosh\left(2r_{A}\right)\right]\sinh\left(2r_{B}\right)\right.\nonumber\\
&&+\left.\left[b+\cosh\left(2r_{B}\right)\right]\sinh\left(2r_{A}\right)\right\}\left(c_{1}^{2}-c_{2}^{2}\right).\nonumber\\
\end{eqnarray}
%%%%%%%%%%%%%%%%%%%%%%%%%%%%%%%%%%%%%%%%%%%%%%%%%%%%%%%%%%%%%%%%%%%%%%%%%%%%%%%%%%%%%%%%%%%%%%%%
The corresponding value of $f$ is denoted by $g(r_{A},r_{B})$.
For a generic two--mode mixed Gaussian state it is again convenient to express the function $g(r_{A},r_{B})$ in terms of $h(r_{A},r_{B})$ as in \eq{gviah}, being
\begin{equation}\label{hh}
h(r_{A},r_{B})=(I_{4}'-I_{3}^{2})/(I_{1}I_{2})\,,
\end{equation}
where
%%%%%%%%%%%%%%%%%%%%%%%%%%%%%%%%%%%%%%%%%%%%%%%%%%%%%%%%%%%%%%%%%%%%%%%%%%%%%%%%%%%%%%%%%%%%%%%%%%%%%%%%%%%%%%
\begin{eqnarray}\label{I123}
I_{1}&=&(a+e^{2r_{A}})(a+e^{-2r_{A}}),\\
I_{2}&=&(b+e^{2r_{B}})(b+e^{-2r_{B}}),\\
I_{3}&=&c_{1}c_{2}.
\end{eqnarray}

Since $h\geq0$ as follows from
the inequality ${\cal I}(A:B)\geq0$, we have to maximize $h$. Introducing new
variables $\lambda=e^{2r_{A}}$ and $\mu=e^{2r_{B}}$, the extremal points of $h$ can be found by
solving the stationarity conditions $\partial h/\partial \lambda=0$ and $\partial h/\partial \mu=0$, respectively,
leading to a set of coupled polynomial equations of the form:
%%%%%%%%%%%%%%%%%%%%%%%%%%%%%%%%%%%%%%%%%%%%%%%%%%%%%%%%%%%%%%%%%%%%%%%%%%%%%%%%%%%%
\begin{eqnarray}\label{extremal}
c_{1}^{2}(a+\lambda)^{2}\mu^{2}+[c_{1}^{2}b(a+\lambda)^{2}-c_{2}^{2}b(a\lambda+1)^{2}\nonumber\\
+c_{1}^{2}c_{2}^{2}a(\lambda^{2}-1)]\mu
-c_{2}^{2}(a\lambda+1)^{2}=0,\nonumber\\
c_{1}^{2}(b+\mu)^{2}\lambda^{2}+[c_{1}^{2}a(b+\mu)^{2}-c_{2}^{2}a(b\mu+1)^{2}\nonumber\\
+c_{1}^{2}c_{2}^{2}b(\mu^{2}-1)]\lambda
-c_{2}^{2}(b\mu+1)^{2}=0.
\end{eqnarray}
%%%%%%%%%%%%%%%%%%%%%%%%%%%%%%%%%%%%%%%%%%%%%%%%%%%%%%%%%%%%%%%%%%%%%%%%%%%%%%%%%%%%
Upon solving the first equation as a quadratic equation with respect to $\mu$ and substituting the
obtained roots into the second equation, one arrives after some algebra at a single
$12^{\rm th}$-order polynomial in the variable $\lambda$ that we do not write here explicitly due
to its complexity. By taking its real roots calculated numerically together
with stationary points on the boundary and picking the one for which $h$ is maximized, we
can finally get the optimal squeezing parameters $r_{A,B}$ of the seed elements $\hat\Pi_{A,B}$ of optimal local POVMs [\eq{POVM}]
 maximizing
the classical mutual information and thus attaining the Gaussian AMID, \eq{GAMID}, of a generic two--mode Gaussian state.
\end{enumerate}

Analytical progress in the calculation of ${\cal A}^G$ can be achieved for special classes of two--mode states, detailed in the following.

\begin{enumerate}
\setcounter{enumi}{3}
\item {\it Symmetric states.}\quad For symmetric states with $a=b$ the maximization of ${\cal I}_{c}^{G}$ can be in principle performed analytically. As the optimal solution is clearly symmetric with $r_{A}=r_{B}\equiv r$, we have only one stationarity condition $dh/d\lambda=0$, where $\lambda=e^{2r}$ is the parameter to optimize over.
After some algebra the stationarity condition boils down to the following fourth-order polynomial equation
%%%%%%%%%%%%%%%%%%%%%%%%%%%%%%%%%%%%%%%%%%%%%%%%%%%%%%%%%%%%%%%%%%%%%%%%%%%%%%%%%%%%
\begin{eqnarray}\label{exsym}
a_{4}\lambda^{4}+a_{3}\lambda^{3}+a_{2}\lambda^{2}+a_{1}\lambda+a_{0}=0,
\end{eqnarray}
%%%%%%%%%%%%%%%%%%%%%%%%%%%%%%%%%%%%%%%%%%%%%%%%%%%%%%%%%%%%%%%%%%%%%%%%%%%%%%%%%%%%
where
%%%%%%%%%%%%%%%%%%%%%%%%%%%%%%%%%%%%%%%%%%%%%%%%%%%%%%%%%%%%%%%%%%%%%%%%%%%%%%%%%%%%
\begin{eqnarray}\label{acoefficients}
a_{0}&=&-c_{2}^{2},\quad a_{1}=-a\left(c_{1}^{2}c_{2}^{2}+3c_{2}^{2}-a^{2}c_{1}^{2}\right),\nonumber\\
a_{4}&=&c_{1}^{2},\quad a_{2}=3a^{2}(c_{1}^{2}-c_{2}^{2}),\nonumber\\
a_{3}&=&a\left(c_{1}^{2}c_{2}^{2}+3c_{1}^{2}-a^{2}c_{2}^{2}\right).
\end{eqnarray}
%%%%%%%%%%%%%%%%%%%%%%%%%%%%%%%%%%%%%%%%%%%%%%%%%%%%%%%%%%%%%%%%%%%%%%%%%%%%%%%%%%%%
The equation (\ref{exsym}) can be solved analytically using the Cardan formulae
but the obtained solutions are rather cumbersome and therefore we do not give them
here explicitly. Calculating the values of the function $h(\lambda)$ in the admissible
real solutions of Eq.~(\ref{exsym}) and also in the stationary points on the
boundary, the point in which the function is maximal gives us the sought optimal squeezing.

\item {\it Squeezed thermal states.}\quad For generally nonsymmetric squeezed thermal states with $c_1=\pm c_2 \equiv c$, the optimal Gaussian POVMs can be derived in a simple closed form by performing the maximization of $h$ in \eq{hh}. We find the following results.
    If the state parameters $a,b$ and $c$ satisfy the inequality
$(a+b+1)^2\geq ab(ab-c^2)$, then the optimality is obtained by homodyne detection ($r_{A,B}\rightarrow\infty$) on both modes giving \eq{ghom}; conversely, if $(a+b+1)^2< ab(ab-c^2)$,
then the optimality is obtained by heterodyne detection ($r_{A}=r_{B}=0$, projection onto coherent states) on both modes
giving
\begin{equation}\label{ghet}
g_{\rm het}=\left[\frac{(a+1)(b+1)}{(a+1)(b+1)-c^2}\right]^{2}.
\end{equation}
Summarizing, the Gaussian AMID of two--mode squeezed thermal states is given by \eq{GAMIDG} with \begin{equation}\label{fgsts}
f=\left\{
    \begin{array}{ll}
      g_{\rm hom}\ \hbox{[\eq{ghom}]}, & (a+b+1)^2\geq ab(ab-c^2); \\
      g_{\rm het}\ \hbox{[\eq{ghet}]}, & \hbox{otherwise.}
    \end{array}
  \right.
  \end{equation}
\item {\it Pure states.}. For pure two--mode squeezed vacuum states  with $a=b=\cosh(2r), c_{1}=-c_{2}=c=\sinh(2r)$, the first case in \eq{fgsts} is always satisfied, and doubly homodyne measurements are optimal for the calculation of the Gaussian AMID. Although this result is quite intuitive (one could guess that in the pure-state case the optimal local Gaussian measurements possessing maximum
Shannon mutual information for distribution of their outcomes would be homodyne detections), the corresponding value of
 ${\cal A}^G$ is strictly bigger than the entropy of entanglement, \eq{EE}, which corresponds to the true AMID ${\cal A}$ globally optimized over joint, possibly non-Gaussian local POVMs as in the definition (\ref{AMID}); and we know from \eq{MIDpuri} that the latter is indeed attained by local photon counting: ${\cal A}^G(\ket{\psi(r)}_{AB})>{\cal A}(\ket{\psi(r)}_{AB})={\cal M}(\ket{\psi(r)}_{AB})=E(\ket{\psi(r)}_{AB})$. Namely,
\begin{equation}\label{GAMIDpuri}
 {\cal A}^G(\ket{\psi(r)}_{AB})) =2{\cal M}(\ket{\psi(r)}_{AB})-\ln[\cosh (2 r)] \,,
 \end{equation}
%\begin{eqnarray}\label{GAMIDpuri}
% {\cal A}^G(\ket{\psi_{AB}(r)}) &=& 4 \cosh ^2(r) \log [\cosh (r)] \\ &-&4 \sinh ^2(r) \ln [\sinh (r)]-\ln    [\cosh (2 r)]\nonumber\,,
% \end{eqnarray}
 which is strictly bigger than the expression in \eq{MIDpuri}.
\end{enumerate}

\begin{figure*}[tbh]
\subfigure{\includegraphics[width=5.7cm]{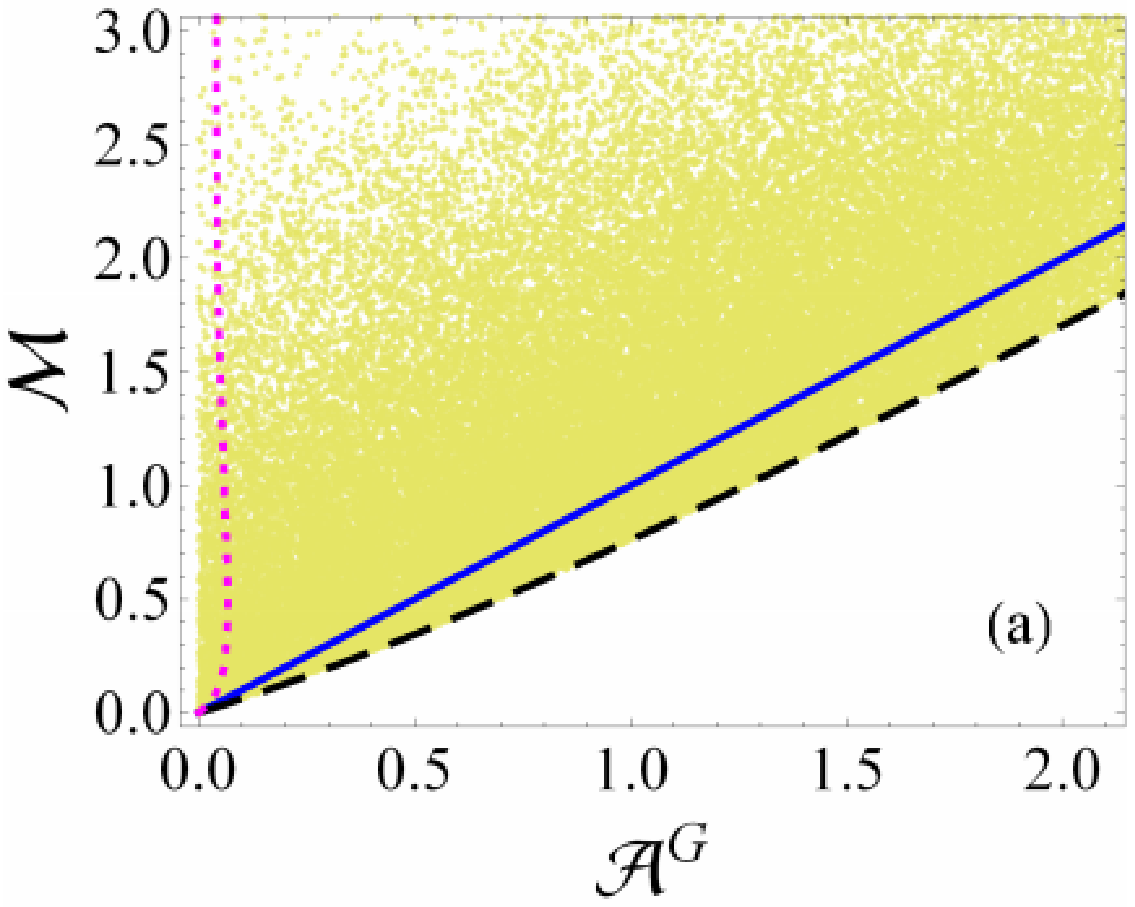}\label{azz}}\hspace*{.2cm}
\subfigure{\includegraphics[width=5.7cm]{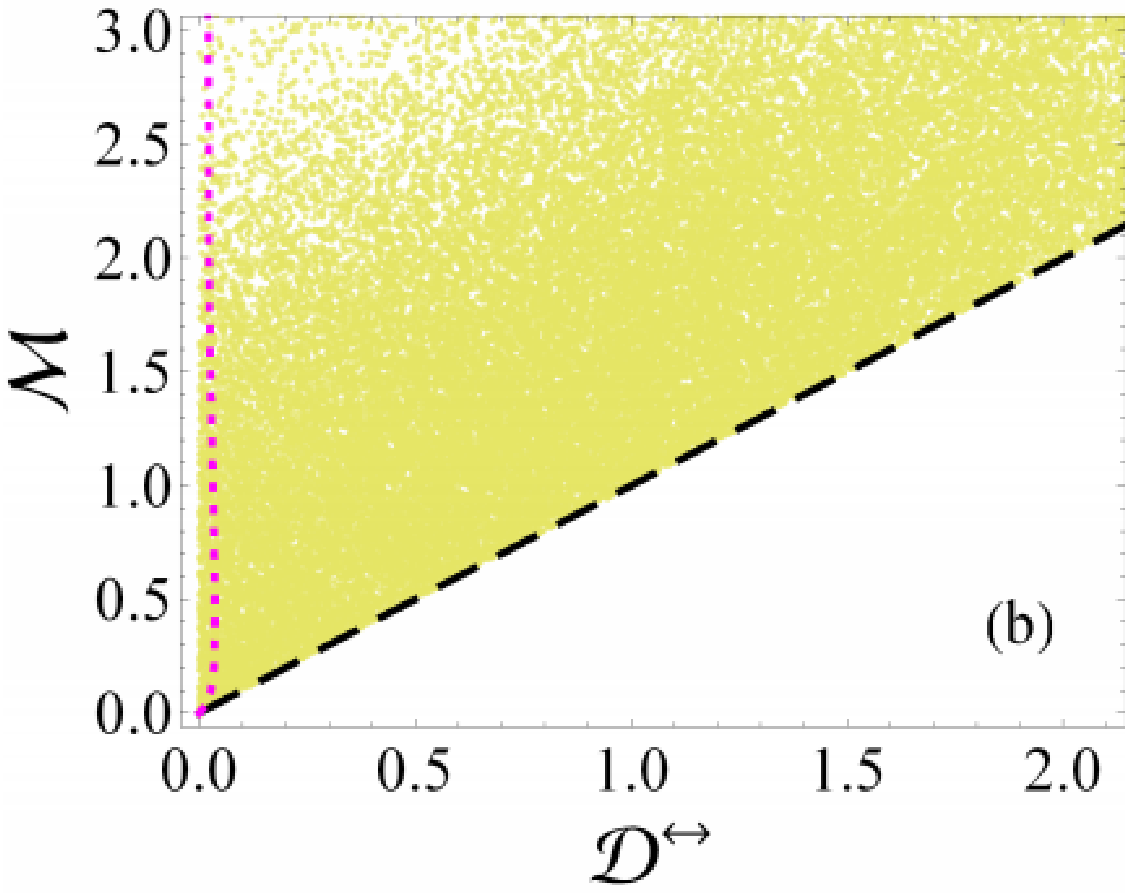}\label{azz2}}\hspace*{.2cm}
\subfigure{\includegraphics[width=5.7cm]{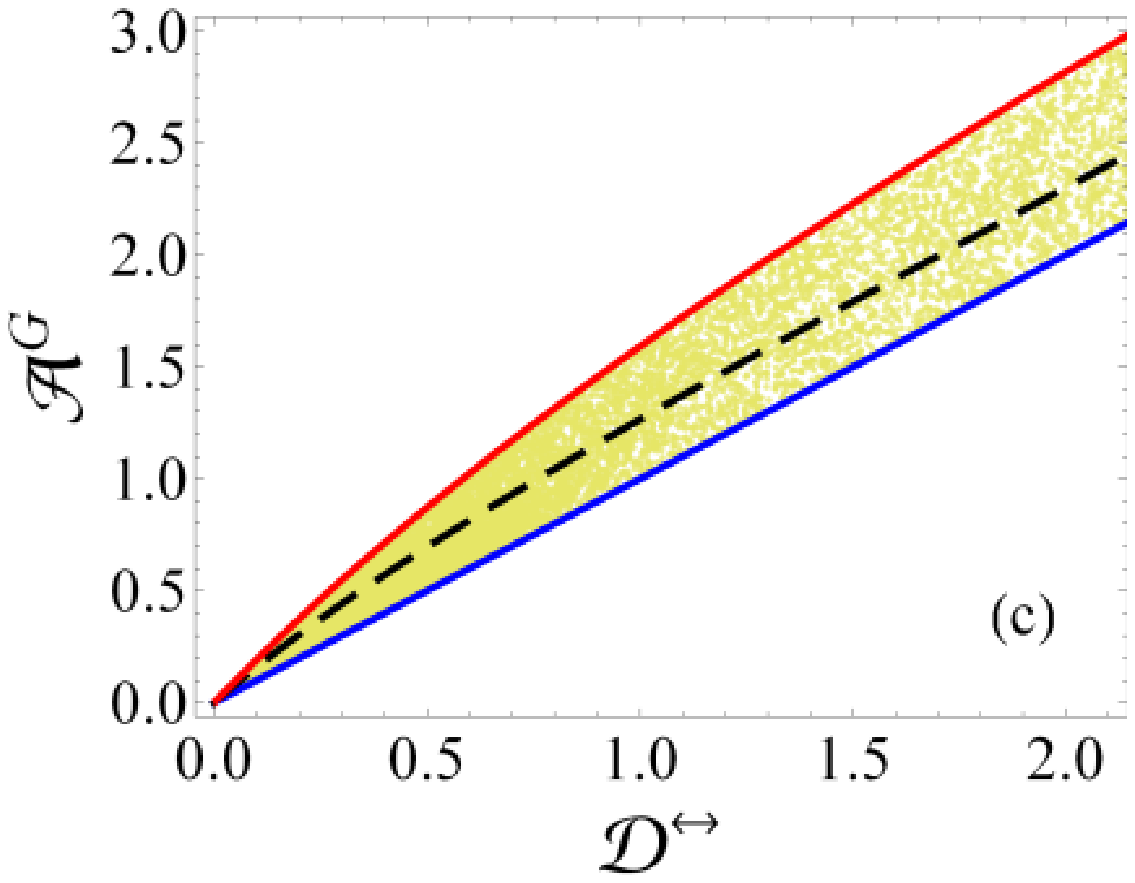}\label{azz3}}
\caption{(Color online) Comparison between (a) MID versus Gaussian AMID, (b) MID versus two-way Gaussian discord, and (c) Gaussian AMID versus two-way Gaussian discord, for $10^5$ randomly generated mixed two--mode Gaussian states. Pure two--mode squeezed states are accommodated on the dashed black curve in all the plots. See text for details of the other boundaries. All the quantities plotted are dimensionless.}
\label{figrandall}
\end{figure*}

 The latter finding entails a novel and interesting showcase in which
 non-Gaussian local measurements on Gaussian states can lead to the extraction of larger correlations than any pair of local Gaussian measurements if quantified by mutual information. In the case of pure two--mode squeezed states this behavior cannot be simply
attributed to the fact that, while the state encodes perfect correlations between photon
numbers (as expressed by $p_{n,m}=0$ for $n\ne m$), there are imperfect
correlations between quadrature operators (as expressed by a nonzero Einstein-Podolsky-Rosen variance
$\langle(\hat x_{A}-\hat x_{B})^{2}\rangle=e^{-2r}$). This is because, while such a variance decays with increasing squeezing $r$, the  gap between the Gaussian AMID ${\cal A}^G$ (obtained upon local homodyne detections) and the true AMID ${\cal A}={\cal M}$ (obtained upon local Fock projections), for pure states, is a monotonically increasing function of $r$ converging to $1-\ln2 \approx 0.3$ in the limit $r \rightarrow \infty$.

A full numerical comparison between Gaussian AMID, MID and Gaussian discord for two--mode Gaussian states will be provided in the next Section.

\section{Comparison between nonclassicality measures for Gaussian states}\label{secDiscuss}

Here we deploy a comprehensive comparative analysis of MID ${\cal M}$ [\eq{MID}, \eq{MIDG}] Gaussian AMID  ${\cal A}^G$ [\eq{GAMID}, \eq{GAMIDG}], and two-way Gaussian quantum discord ${\cal D}^\leftrightarrow$ [\eq{twoway}, Ref.~\cite{AD_10}], as tools to quantify the quantumness of correlations in arbitrary two--mode Gaussian states via entropic descriptions of the state disturbance following suitable local measurements on one or both local parties. The results of the previous Sections show that in \ineq{boundamid}, either quantity on the right hand side can be the smallest on particular instances of two--mode Gaussian states, suggesting that the subset of two--mode Gaussian states whose true AMID $\cal A$, \eq{AMID}, is necessarily optimized by non-Gaussian measurements, might have a finite volume in the space of general two--mode Gaussian states.

To confirm this interesting feature and to investigate the tightness of the hierarchy established in \eq{hierardy}, we have generated a large number of random two--mode Gaussian states (up to $10^6$), and for each of them we have evaluated the three symmetric nonclassicality indicators ${\cal D}^\leftrightarrow$, ${\cal A}^G$, and ${\cal M}$, following the prescriptions of the preceding Sections. The resulting analysis is illustrated in Fig.~\ref{figrandall}.
Panel (a) shows that, while the MID can be arbitrarily larger than the Gaussian AMID in principle, there is nonetheless a finite region in the (${\cal A}^G$,  ${\cal M}$) diagram that allocates two--mode Gaussian states for which even non-optimized non-Gaussian measurements (specifically, photon counting) result in a larger classical mutual information, hence minimize the quantum correlations in the definition of the AMID, compared to the optimal Gaussian POVMs. In this study we did not consider other non-Gaussian measurements apart from photon counting (that enters in the definition of MID because Gaussian mixed states are essentially thermal states): Therefore, we can expect that the region in which general non-Gaussian measurements are optimal for the AMID can be in principle much larger than the one highlighted by the present study (that is located below the blue online segment of equation ${\cal A}^G={\cal M}$ in Fig.~\ref{azz}). Still, our finding is perhaps one of the most striking instances of an operational quantum informational measure for Gaussian states that can gain a significant optimization by the use of suitable non-Gaussian operations. Non-Gaussian operations can sometimes reveal quantumness more accurately, thus unleashing more precisely the available nonclassical resources, than the best Gaussian measurements, on certain two--mode Gaussian states, including quite remarkably all two--mode pure Gaussian states. Notice however that the rigidity in choosing the non-Gaussian measurements for the evaluation of MID, excluding any optimization procedure, results in most of the cases into a very loose overestimation of the quantum correlations, as testified by the unbounded region above the straight blue line, filled by states where certainly joint photon counting is not optimal for the AMID.
Interestingly, pure states embody the lower bound (dashed black curve) in Fig.~\ref{azz}: they are therefore the states where the Gaussian AMID realizes the most dramatic overestimation of the true AMID, that nonetheless can never exceed $\approx 0.3$ as computed in the previous Section. A family of states sitting on the blue line in Fig.~\ref{azz} will be characterized shortly.

\begin{figure*}[tbh]
\includegraphics[width=17cm]{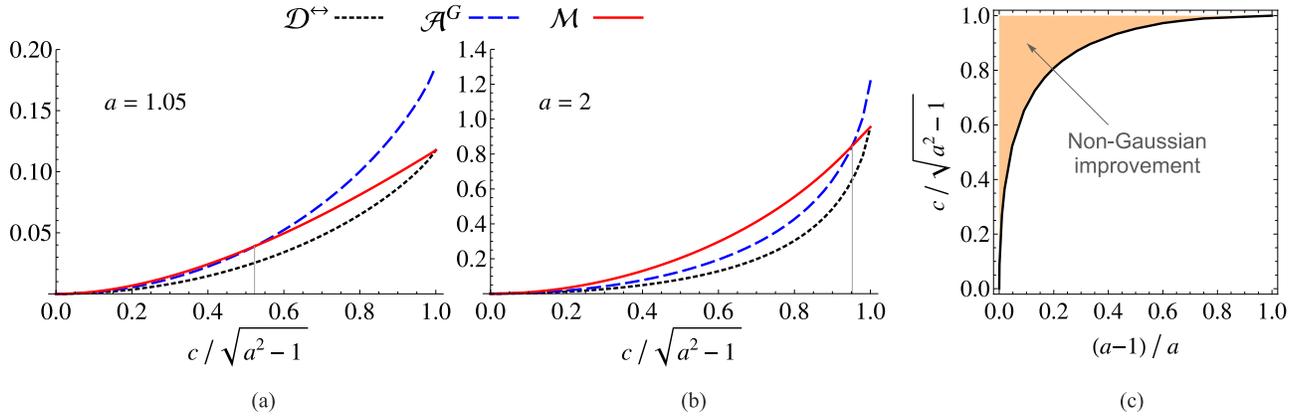}
\caption{(Color online) Comparison of different measures of quantum correlations for two--mode symmetric squeezed thermal Gaussian states ($b=a, c_1=-c_2=c$). Panels (a)-(b): Two-way Gaussian quantum discord ${\cal D}^{\leftrightarrow} = \max\{{\cal D}^{\leftarrow},{\cal D}^{\rightarrow}\}$ (dotted black line), Gaussian AMID ${\cal A}^{G}$ associated to optimal bi-local Gaussian POVMs (dashed blue line), and unoptimized MID ${\cal M}$ associated to joint photon counting (solid red line), plotted versus the normalized state covariance parameter $c/\sqrt{a^2-1}$, for (a) $a=1.05$ and (b) $a=2$. The AMID ${\cal A}$ optimized over all possible (Gaussian and non-Gaussian) measurements is certainly ${\cal A} \le \min\{{\cal M}, {\cal A}^G\}$. Both MID and Gaussian AMID majorize the Gaussian discord, but for $c$ bigger than a certain threshold value $c^{\star}(a)$ (ticked by a vertical gray line in the plots) one has ${\cal M} < {\cal A}^{G}$, meaning that non-Gaussian measurements become necessarily optimal for the AMID. Panel (c) depicts the threshold curve $c^{\star}(a)$ (solid black line), defined by the condition ${\cal M} = {\cal A}^{G}$, in the normalized parameter space $\{(a-1)/a, c/\sqrt{a^2-1}\}$. The shaded (orange) region above the threshold line allocates instances of the considered family of states, lying in the neighborhood of pure two--mode squeezed states, where certainly Gaussian POVMs are not globally optimal for the AMID, since photon counting results in a lower figure of merit.  Below the threshold, either Gaussian measurements are optimal or there may exist some more general non-Gaussian measurement that  achieves the absolute minimum in ${\cal A}$: our analysis cannot rule out this possibility. All the quantities plotted are dimensionless.}  \label{figcross}
\end{figure*}

Before that, let us comment on the other panels of Fig.~\ref{figrandall}. Panel (b) shows as expected, and in full analogy with the case of two qubits \cite{amid}, that in general the unoptimized MID based on photon counting is a very loose upper bound to quantum discord for two--mode Gaussian states (reducing to it on pure states, depicted as dashed black again), unbounded from above and relentlessly approaching arbitrarily large values even for states with nearly vanishing quantum correlations as quantified by the (Gaussian) discord. This should discourage the usage of MID in general as it almost always provides overestimations, rather than reliable quantifications, of nonclassicality of bipartite correlations.

The last panel shows also a somehow analogous situation to the two--qubit case \cite{amid}: the Gaussian AMID is intimately related to discord, and admits upper and lower bounds at a given value of the two-way Gaussian discord. The lower (blue online) boundary in panel (c) accommodates states for which the two quantifiers give identical prescriptions for measuring quantum correlations. These are states with CM in standard form, \eq{gamma}, given by
\begin{equation}
\label{cmivette}
\begin{split}
&a=\cosh(2s),\,b=\cosh^2(r)\cosh(2s) + \sinh^2(r),\,\\
&c_{1}=-c_{2}=\cosh r \sinh(2s)\,,
\end{split}
\end{equation}
in the limit $r \rightarrow \infty$. They are characterized by  ${\cal A}^G={\cal D}^\leftrightarrow=2 \sinh^2 s\ln(\coth s)$. Pure states fill once more the dashed black curve, for which ${\cal D}^\leftrightarrow={\cal E}$, \eq{MIDpuri}, but ${\cal A}^G$ is strictly bigger, \eq{GAMIDpuri}. The upper (red online) boundary in Fig.~\ref{azz3} can be spanned for instance by symmetric squeezed thermal states ($b=a$, $c_1=-c_2=c$), with $a \gg 1$ and $c \in [0, \sqrt{a^2-1})$. Upper and lower boundaries ideally conjoin asymptotically for diverging discord and Gaussian AMID.

We can now analyze in detail the competition between the MID associated to photon counting (typically very loose, but optimal on pure states) and the Gaussian AMID (very accurate for mixed and strongly correlated Gaussian states) to maximize the classical mutual information, hence minimizing the AMID, \eq{boundamid}, on two--mode Gaussian states. We believe it be relevant to focus on the  class of two--mode symmetric squeezed thermal Gaussian states ($b=a, c_1=-c_2=c$), for which the involved measures can be simply evaluated \footnote{For squeezed thermal states, the Gaussian discord is optimized by a local heterodyne detection \cite{Giorda_10,AD_10}.}. We plot in Fig.~\ref{figcross} [(a),(b)] a comparison of the three quantumness measures studied in this work as a function of the rescaled state parameters $a$ and $c$. We see that there is a certain threshold value $c^{\star}(a)$ beyond which the Gaussian POVMs are no longer optimal for the AMID, and non-Gaussian measurements such as photon counting (via MID) provide a more accurate result, culminating in the extreme case of pure states where those specific measurements are globally optimal. Panel (c) depicts the threshold in the parameter space, highlighting the region where our analysis conclusively reveals the necessity of non-Gaussian measurements for the global optimization of AMID and classical mutual information of the considered class of two--mode Gaussian states. As previously remarked, this region can be in principle (and is likely to be so) much larger. Yet, it certainly occupies a finite volume in the space of general two--mode Gaussian states. Notice that for all the Gaussian states in such a region, non-Gaussian measurements allow one to extract stronger  correlated measurement records compared to any bi-local Gaussian measurement, as the classical mutual information is maximized by non-Gaussian detections. The states attaining the threshold identified in this analysis, are an instance of states filling up the blue line in Fig.~\ref{azz}.

\begin{figure}[tb]
\includegraphics[width=8cm]{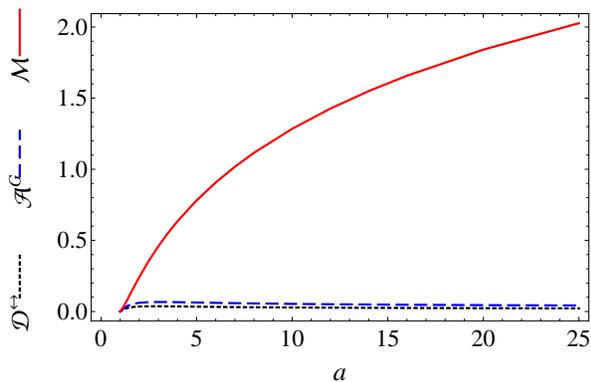}
\caption{(Color online) Plot of MID (solid red line), Gaussian AMID (dashed blue line) and Gaussian two-way quantum discord (dotted black line) as a function of the parameter $a$ for the two--mode Gaussian states of \eq{statistrani}. These curves give origin to the dotted magenta lines in Fig.~\ref{figrandall}(a),(b). All the quantities plotted are dimensionless.}  \label{figstrani}
\end{figure}

Finally, we exhibit an example  family of two--mode Gaussian states where, on the opposite end, the MID based on non-Gaussian detections is a highly inaccurate measure of quantum correlations. These states sit on the quasi-vertical dashed (magenta online) curves in Fig.~\ref{figrandall}(a) and (b). They are symmetric states with \begin{equation}\label{statistrani}
\mbox{$b=a$, $c_2=0$, and $c_1 = (a^2-1-\ln a)/a$}.
 \end{equation}
 As apparent from Fig.~\ref{figstrani}, their Gaussian discord and Gaussian AMID stay limited (smaller than $\approx 0.06$) and rigorously vanish in the asymptotic limit $a \rightarrow \infty$. On the other hand, their MID arising from Fock projections increases arbitrarily and diverges for $a \rightarrow \infty$, embodying an extreme overestimation of some vanishing quantum correlations. Clearly there will be many more families of Gaussian states where such a behavior will arise.

\section{Nonclassicality versus entanglement}\label{secEnt}

Here we present a numerical comparison between nonclassicality of correlations, measured by means of the Gaussian AMID [\eq{GAMID}], and entanglement, quantified by the Gaussian EoF [\eq{geof}] \cite{geof}, for generally mixed two--mode Gaussian states. A similar analysis was performed in Ref.~\cite{AD_10}, with (Gaussian) discord used as a nonclassicality indicator.

\begin{figure}[bt]
\includegraphics[width=8cm]{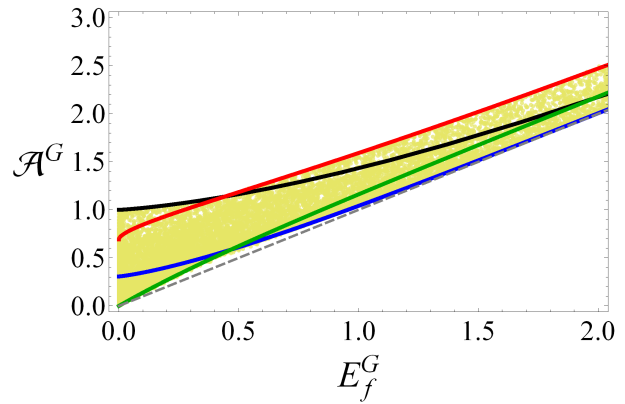}
\caption{(Color online) Plot of Gaussian AMID ${\cal A}^G$ versus Gaussian EoF $E_f^G$ for $10^5$ random two--mode Gaussian states. The dashed line of equation ${\cal A}^G=E_f^G$ stands as a lower bound for the physically admitted region. Refer to the main text for details of the other curves. All the quantities plotted are dimensionless.}\label{figrefent2d}
\end{figure}

Fig.~\ref{figrefent2d} shows the distribution of Gaussian AMID versus Gaussian EoF for a sample of $10^5$ randomly generated two--mode Gaussian states. In analogy with the case of Gaussian discord vs Gaussian EoF, it is possible to identify upper and lower bounds on the Gaussian AMID ${\cal A}^G$ at fixed entanglement $E_f^G$. Interestingly, our numerical exploration shows that for all two--mode Gaussian states $\hat{\rho}_{AB}$, it is
\begin{equation}\label{majoriz}
{\cal A}^G(\hat{\rho}_{AB}) \ge E_f^G(\hat{\rho}_{AB})\,.
\end{equation}
This provides a novel hierarchical relationship between different types of nonclassical resources, entanglement $E_f^G$, and more general measurement-induced quantum correlations ${\cal A}^G$: On the basis of the hereby employed measures (both  symmetric by construction and restricted to a fully Gaussian scenario), the latters appear to always encompass and exceed entanglement itself for two--mode generally mixed Gaussian states. A similar relationship does not hold for discord, which can be smaller as well as larger than entanglement of formation, even in a Gaussian scenario \cite{AD_10,James}.

We can provide two families of two--mode Gaussian states for which \eq{majoriz} becomes asymptotically tight.
One such class is provided, e.g., by symmetric squeezed thermal states, whose standard form CM is as in \eq{gamma} with
\begin{equation}\label{gmems}
b=a\,,\quad c_1=-c_2=a-\tilde{\nu}\,\,,\end{equation}
where
\begin{equation}\label{grange}
\tilde{\nu}>0\,,\quad a \ge \max\{\tilde{\nu},\,(1+\tilde{\nu}^2)/(2\tilde{\nu})\}\,.
\end{equation}
The Gaussian EoF of these states (equal to the true EoF minimized over all possible pure-state decompositions, by virtue of the symmetry of the states \cite{giedke03}) is a simple monotonically decreasing function of the positive parameter $\tilde{\nu}$,
\begin{equation}
\label{eofsym}
E_f^G(\tilde{\nu})=\frac{(1+\tilde{\nu })^2 \ln\left[\frac{(1+\tilde{\nu })^2}{4 \tilde{\nu }}\right]-(1-\tilde{\nu })^2 \ln\left[\frac{(1-\tilde{\nu })^2}{4 \tilde{\nu }}\right]}{4 \tilde{\nu }}\,,
\end{equation}
if $\tilde{\nu}<1$, and $E_f^G(\tilde{\nu})=0$ otherwise.
The Gaussian AMID can be computed analytically according to the prescription of Sec.~\ref{secGAMID}, and for any fixed $\tilde{\nu}$ (i.e., fixed Gaussian EoF) one can find the optimal value of the parameter $a$, in the range defined by \eq{grange}, that minimizes ${\cal A}^G$. The resulting ${\cal A}^G$ as a function of $E_f^G$ is plotted in Fig.~\ref{figrefent2d} as a green curve: For this family of states, the Gaussian AMID approaches the Gaussian EoF as the latter tends to zero, tending to saturate  Ineq.~(\ref{majoriz}) asymptotically in the regime of infinitesimal correlations.

Furthermore, let us consider another class of symmetric two--mode states, whose standard form CM is as in \eq{gamma} with
\begin{equation}\label{glems}
b=a,\,c_1 = a-(1+\tilde{\nu}^2)/(2a),\,c_2=a-(2a)/(1+\tilde{\nu}^2)\,,
 \end{equation}
where the parameter range is the same as in \eq{grange}.
 The Gaussian EoF of these states is still given by \eq{eofsym}, while in the limit $a \rightarrow \infty$ one can show that their Gaussian AMID is attained by homodyne detections on both modes, yielding   ${\cal A}^G(\tilde{\nu})=1-\ln(4\tilde{\nu})+\ln(1+\tilde{\nu}^2)$. The corresponding Gaussian AMID vs Gaussian EoF curve for these states is depicted in Fig.~\ref{figrefent2d} as a blue curve: In this case, the Gaussian AMID approaches the Gaussian EoF as the latter tends to infinity, also saturating  Ineq.~(\ref{majoriz}) asymptotically.

It has to be underlined that the two presented Gaussian families are just examples to show that the bound in \eq{majoriz} can be asymptotically tight, and we remark that the combination of the two presented curves does not provide a strict lower bound to Gaussian AMID against Gaussian EoF for all two--mode Gaussian states, as it is evident from the presence of some random points below the intersection of the two (green and blue) curves  -- however above the dashed line corresponding to the bound of \eq{majoriz} [see Fig.~\ref{figrefent2d}].

On the other hand, a tight upper bound on the Gaussian AMID at fixed Gaussian EOF can be identified. We found it numerically to be constituted by the maximum of two branches,
\begin{equation}\label{gubound}
 {\cal A}^G(\hat{\rho}_{AB}) \le
\left\{
  \begin{array}{l}
    1 + 2 \ln(1+\tilde{\nu})-\ln(4\tilde{\nu}),  \\
    \quad   4/{\rm e} -1 \le \tilde{\nu} \le 1; \\
        \\
    \ln(1+\tilde{\nu})-\ln(\tilde{\nu}),   \\
     \quad 0 < \tilde{\nu} < 4/{\rm e} -1;
  \end{array}
\right.
\end{equation}
where $E_f^G(\hat{\rho}_{AB})$ is given by \eq{eofsym}.

The first expression in \eq{gubound} corresponds to the Gaussian AMID of the states of \eq{cmivette} with $s \rightarrow \infty$ and $r = 2 \tanh^{-1}(\sqrt{\tilde{\nu}})$, and is depicted as a black curve in Fig.~\ref{figrefent2d}. It provides an upper bound for all two--mode Gaussian states distributed in the $\{E_f^G, {\cal A}^G\}$ plane, in the region of moderate entanglement; such a bound converges to 1 in the separability limit  ($\tilde{\nu} = 1$). We can thus conclude that the degree of nonclassicality of correlations in separable Gaussian states is always nonzero (apart from the trivial case of uncorrelated, product states) but stays nevertheless limited: It can at most reach unity, whether measured by the  discord \cite{Giorda_10,AD_10} or by the Gaussian AMID.

The second expression in \eq{gubound} corresponds instead to the Gaussian AMID of the states of \eq{gmems} in the limit $a \rightarrow \infty$. It bounds from above the value of ${\cal A}^G$ for all two--mode Gaussian states with fixed $E_f^G \gtrapprox 0.441$ [where this number is obtained by setting $\tilde{\nu} =  4/{\rm e} -1$ in \eq{eofsym}]. In the limit of infinite entanglement ($\tilde{\nu} \rightarrow 0$), the upper bound on the Gaussian AMID converges to $E_f^G + \ln 4 - 1$. Combining this observation with the lower bound (\ref{majoriz}), we have that, interestingly, the following sandwich relation holds for all two--mode Gaussian states with $E_f^G \gg 0$\,,
\begin{equation}\label{panino}
 E_f^G(\hat{\rho}_{AB}) \le {\cal A}^G(\hat{\rho}_{AB})  \le  E_f^G(\hat{\rho}_{AB}) + \ln4 -1\,.
\end{equation}

In the previous analysis, we have identified some similarities as well as some key differences in the quantification of nonclassical correlations versus entanglement of Gaussian states when Gaussian AMID rather than quantum discord are employed. In order to have a visual comparison between the two nonclassicality indicators and the Gaussian EoF, we focus on the relevant two-parameter class of symmetric squeezed thermal states $\hat{\rho}_{AB}^{sts}$, with CM as in \eq{gmems}. For these states, the entanglement is given by \eq{eofsym} (independently of $a$), while the discord can be written as \cite{Giorda_10,AD_10}
\begin{equation}\label{stsDis}
\begin{split}
{\cal D}^\leftarrow & (\hat{\rho}_{AB}^{sts})
\\ =& \frac{1}{2(1+a)}\Bigg\{\left(4 a (\tilde{\nu} +1)-2 \tilde{\nu} ^2\right) \tanh ^{-1}\left(\frac{a+1}{2 a \tilde{\nu} +a-\tilde{\nu} ^2}\right)\\ &-4 (a+1) \sqrt{\tilde{\nu}  (2 a-\tilde{\nu} )} \tanh ^{-1}\left(\frac{1}{\sqrt{\tilde{\nu}  (2 a-\tilde{\nu} )}}\right) \\
&+  a^2 \ln \left(\frac{a+1}{a-1}\right)-\ln \left[\frac{(a+1) \left(2 a \tilde{\nu} -\tilde{\nu} ^2-1\right)}{(a-1) (\tilde{\nu} +1) (2 a-\tilde{\nu} +1)}\right]\! \Bigg\},
 \end{split}
\end{equation}
and the Gaussian AMID reads
\begin{equation}\label{stsGam}
\begin{split}
&{\cal A}^G  (\hat{\rho}_{AB}^{sts}) =
-\ln \left(\frac{2 a \tilde{\nu} -\tilde{\nu} ^2-1}{a^2-1}\right)+2 a \coth ^{-1}(a) \\
& -2 \sqrt{\tilde{\nu}  (2 a-\tilde{\nu} )} \tanh ^{-1}\left(\frac{1}{\sqrt{\tilde{\nu}  (2 a-\tilde{\nu} )}}\right) \\
&-\ln
\left\{
       \begin{array}{ll}
         \frac{a}{\sqrt{a^2-(a-\tilde{\nu} )^2}}, & 1+a \left(4+a \left(4-2 a \tilde{\nu} +\tilde{\nu} ^2\right)\right)\geq 0; \\
         \frac{(a+1)^2}{(a+1)^2-(a-\tilde{\nu} )^2}, & \hbox{otherwise.}
       \end{array}
     \right.
\end{split}
\end{equation}

\begin{figure}[tb]
\includegraphics[width=8cm]{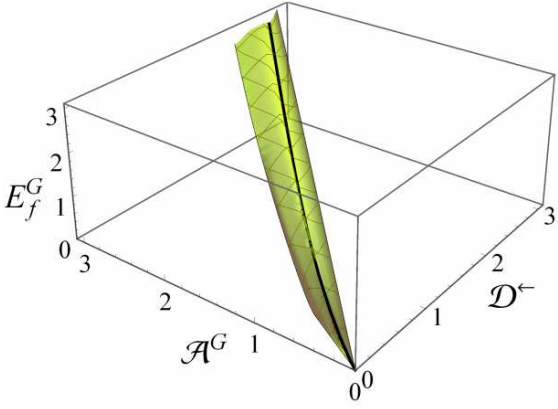}
\caption{(Color online) 3D Plot of Gaussian EoF $E_f^G$ [\eq{eofsym}] versus Gaussian discord ${\cal D}^\leftarrow$ [\eq{stsDis}] and Gaussian AMID ${\cal A}^G$ [\eq{stsGam}] for two--mode symmetric squeezed thermal Gaussian states, characterized by their covariance parameters $a$ and $\tilde{\nu}$ [see \eq{gmems}]. The solid black line accommodates pure states ($\tilde{\nu} = a - \sqrt{a^2-1}$).
All the quantities plotted are dimensionless.}\label{fig3d}
\end{figure}

Fig.~\ref{fig3d} shows $E_f^G$ plotted versus ${\cal D}^{\leftarrow}$ and ${\cal A}^G$ for this particular class of Gaussian states, spanned by $a$ and $\tilde{\nu}$. All two--mode symmetric squeezed thermal states sit on a two-dimensional surface in the space of the three entropic nonclassicality indicators, providing a direct evidence of the intimate yet intricate relationship between the different aspects of quantumness in Gaussian states. One can notice the branch of separable states, in the plane $E_f^G=0$ with generally nonzero discord and Gaussian AMID; while for all entangled squeezed thermal states, $E_f^G$ can be exactly recast as  a function of ${\cal D}^{\leftrightarrow}$ and ${\cal A}^G$: Knowledge of two nonclassicality quantifiers fixes the third one univocally. We remark that such a strict result does not extend to more general two--mode Gaussian states, which distribute filling a more complex, finite-volume three-dimensional region in the space $\{{\cal D}^{\leftarrow},{\cal A}^G, E_f^G\}$. Figures \ref{azz3} and \ref{figrefent2d} of this paper, and Figure 1(right) of Ref.~\cite{AD_10}, represent the two-dimensional projections of such a region onto the planes $\{{\cal D}^{\leftarrow},{\cal A}^G\}$, $\{E_f^G,{\cal A}^G\}$, and $\{E_f^G,{\cal D}^\leftarrow\}$, respectively.

\section{Conclusions}\label{secConcl}

We have performed an exhaustive study of nonclassical correlations in generic two--mode Gaussian states using
information-theoretic nonclassicality quantifiers, in particular the `measurement-induced disturbance' (MID) and its Gaussian-optimized
version, the Gaussian `ameliorated measurement-induced disturbance' (Gaussian AMID). For a given Gaussian state the MID
is a gap between its quantum mutual information --- quantifying the total correlations --- and the classical mutual
information of outcomes of local Fock-state detections (detections of local eigenprojectors) --- that captures a specific type of
non-Gaussian classical correlations in the state. The Gaussian AMID is, on the other hand, a gap between the quantum mutual
information and the maximal classical mutual information that can be obtained by local Gaussian measurements, the latter quantifying the
maximum classical correlations that can be extracted from the state by local Gaussian processing. An analytical form of the Gaussian
AMID can be derived for the important subclasses of symmetric states and squeezed thermal states which include pure states, while for a
generic mixed Gaussian state one has to find roots of a higher-order polynomial in a single variable, which can be solved efficiently
by numerical means. Further analysis reveals that MID is mostly larger than Gaussian AMID and therefore overestimates the amount of
nonclassical correlations. In fact, for a fixed value of Gaussian AMID, it is possible to find states with an arbitrarily large MID, even if the Gaussian AMID is infinitesimally small. On the other hand, there also exists a volume of Gaussian states encompassing pure states for
which the MID is strictly smaller than the Gaussian AMID, which surprisingly unveils the importance of non-Gaussian measurements
for the correct assessment of the amount of nonclassical correlations in Gaussian states.

We have further compared the MID and
Gaussian AMID with the two-way Gaussian discord. We found that again there is no upper bound on MID for a fixed value of discord
but a close upper bound (as well as a lower one) does exist for the Gaussian AMID. Finally, we have also compared the Gaussian AMID with the Gaussian entanglement of formation, identifying lower and upper bounds for the former as a function of the latter. In particular, the Gaussian AMID turns out to always exceed the Gaussian entanglement of formation for all two--mode Gaussian states, enforcing a novel hierarchy between two different forms of nonclassicality. Exact relations between Gaussian AMID, Gaussian discord and Gaussian entanglement of formation can be formulated for special families of Gaussian states, such as the symmetric squeezed thermal states.

On a more technical side, we have shown that symmetric measures of (non)classicality of correlations such as the classical mutual information and the Gaussian AMID are tractable by analytical tools
for important subclasses of states even considering optimization
over local generalized Gaussian measurements. Our results also
demonstrate that non-Gaussian
processing for correct quantification of (non)classical correlations
in Gaussian states is in order.

We believe that these findings
will inspire further research on the characterization of quantum correlations in the
Gaussian scenario and beyond.

\acknowledgments

We are grateful to A. Datta and M. Paternostro for very fruitful discussions on the topics of this paper.
L. M., R. T., and N. K. acknowledge the EU grant under FET-Open project COMPAS
(212008). L. M.  has been supported by projects ``Measurement and Information in Optics,'' (MSM 6198959213) and
Center of Modern Optics (LC06007) of the Czech Ministry of Education and
the project of GACR No. 202/08/0224. R. T. and N. K. are grateful for the support from SUPA (Scottish  Universities Physics Alliance).

\appendix
\section{Reduction to covariant rank-one POVMs}\label{secAppPOVM}

Here we prove that covariant rank-one POVMs, i.e. POVMs of the form \eq{POVM} with a pure seed state $\hat\Pi_j$,
are optimal among all Gaussian POVMs
for the evaluation of the classical mutual information or, equivalently, of the Gaussian AMID, \eq{GAMID}.

The measurement of local POVMs (\ref{POVM}) on a Gaussian state $\hat{\rho}_{AB}$ gives the outcome
$d=(d_{A}^{T},d_{B}^{T})^{T}$ distributed according to
%%%%%%%%%%%%%%%%%%%%%%%%%%%%%%%%%%%%%%%%%%%%%%%%%%%%%%%%%%%%%%%%%%%%%%%%%%%%%%%%%%%%%%%%%%%%%%%%%%%%%%%%%
\begin{equation}\label{Pd1}
P(d)=\mbox{Tr}[\hat{\Pi}_{A}(d_{A})\otimes\hat{\Pi}_{B}(d_{B})\hat{\rho}_{AB}].
\end{equation}
%%%%%%%%%%%%%%%%%%%%%%%%%%%%%%%%%%%%%%%%%%%%%%%%%%%%%%%%%%%%%%%%%%%%%%%%%%%%%%%%%%%%%%%%%%%%%%%%%%%%%%%%%
Making use of the overlap formula for Wigner functions \cite{Leonhardt_97},
the distribution can be expressed as
%%%%%%%%%%%%%%%%%%%%%%%%%%%%%%%%%%%%%%%%%%%%%%%%%%%%%%%%%%%%%%%%%%%%%%%%%%%%%%%%%%%%%%%%%%%%%%%%%%%%%%%%%
\begin{eqnarray}\label{Pd2}
P(d)&=&(2\pi)^{2}\int W_{\hat{\Pi}_{A}(d_{A})}(r_{A})W_{\hat{\Pi}_{B}(d_{B})}(r_{B})\nonumber\\
&&\times W_{\hat{\rho}_{AB}}(r_{A},r_{B}){\rm d}^{2}r_{A}{\rm d}^{2}r_{B}.
\end{eqnarray}
%%%%%%%%%%%%%%%%%%%%%%%%%%%%%%%%%%%%%%%%%%%%%%%%%%%%%%%%%%%%%%%%%%%%%%%%%%%%%%%%%%%%%%%%%%%%%%%%%%%%%%%%%
Substituting into the formula from Eq.~(\ref{W}), where we set $r=(r_{A}^{T},r_{B}^{T})^{T}$, and performing the integration,
we obtain the distribution (\ref{Pd1}) in the form:
%%%%%%%%%%%%%%%%%%%%%%%%%%%%%%%%%%%%%%%%%%%%%%%%%%%%%%%%%%%%%%%%%%%%%%%%%%%%%%%%%%%%%%%%%%%%%%%%%%%%%%%%%%
\begin{equation}\label{Pd}
P(d)=\frac{1}{\pi^2\sqrt{\det(\gamma+\gamma_{A}\oplus\gamma_{B}})}e^{-d^{T}(\gamma+\gamma_{A}\oplus\gamma_{B})^{-1}d},
\end{equation}
%%%%%%%%%%%%%%%%%%%%%%%%%%%%%%%%%%%%%%%%%%%%%%%%%%%%%%%%%%%%%%%%%%%%%%%%%%%%%%%%%%%%%%%%%%%%%%%%%%%%%%%%%%
where $\gamma_{A,B}$ are CMs of the seed elements of POVMs (\ref{POVM}) and $\gamma$ is the CM of the state $\hat{\rho}_{AB}$.
The CMs $\gamma_{A,B}$ can be expressed as $\gamma_{j}=\gamma_{j}^{(\pi)}+N_{j}$, where $\gamma_{j}^{(\pi)}=S_{j}^{-1}(S_{j}^{T})^{-1}$
is a pure-state CM ($S_{j}$ symplectically diagonalizes $\gamma_{j}$) and $N_{j}=(\nu_{j}-1)\gamma_{j}^{(\pi)}$ is a positive-semidefinite matrix ($\nu_{j}\geq1$ is a symplectic eigenvalue of $\gamma_{j}$).
%This enables us to express the distribution (\ref{Pd}) as the convolution
%%%%%%%%%%%%%%%%%%%%%%%%%%%%%%%%%%%%%%%%%%%%%%%%%%%%%%%%%%%%%%%%%%%%%%%%%%%%%%%%%%%%%%%%%%%%%%%%%%%%%%%%%%
%\begin{equation}\label{convolution}
%P(d)=\int P_{A}(d_{A}-d_{A}')P_{B}(d_{B}-d_{B}')P^{(p)}(d_{A}',d_{B}')d^{2}d_{A}'d^{2}d_{B}',
%\end{equation}
%%%%%%%%%%%%%%%%%%%%%%%%%%%%%%%%%%%%%%%%%%%%%%%%%%%%%%%%%%%%%%%%%%%%%%%%%%%%%%%%%%%%%%%%%%%%%%%%%%%%%%%%%%
 Therefore, the outcome $d_{j}$, $j=A,B$ of a generic POVM $\hat{\Pi}_j(d_j)$, with the seed element
$\hat{\Pi}_j$ being a mixed state with CM $\gamma_{j}$, can be expressed as $d_{j}=d_{j}^{(\pi)}+\chi_{j}$, where $d_{j}^{(\pi)}$ is the
outcome of POVM with pure-state seed element having CM $\gamma_{j}^{(\pi)}$ and
$\chi_{A},\chi_{B}$ are mutually uncorrelated random variables
uncorrelated with $d_{A}^{(\pi)},d_{B}^{(\pi)}$ obeying Gaussian distributions with classical correlation matrices
$N_{A}$ and $N_{B}$, respectively. Since such processing of variables $d_{j}^{(\pi)}$ cannot increase their
Shannon mutual information due to what is now known as data processing inequality (derived first for continuous
random variables in \cite{Kolmogorov_56}), we can restrict  without loss of generality to optimization over projections
onto pure states.

Note, that covariant measurements (\ref{POVM}) with pure-state seed elements maximize classical mutual information
even within the framework of a larger class of generally noncovariant Gaussian POVMs possessing the structure \cite{Mista_08,Giorda_10}
%%%%%%%%%%%%%%%%%%%%%%%%%%%%%%%%%%%%%%%%%%%%%%%%%%%%%%%%%%%%%%%%%%%%%%%%%%%%%%%%%%%%%%%%%%%%%%%%%%%%%%%%%
\begin{eqnarray}\label{noncovariant}
\hat{\pi}_{j}(z_{j})=\frac{p_{j}(y_{j})}{2\pi} \hat{D}_j(d_j)\hat{\Pi}_j(y_{j}) \hat{D}_j^\dagger(d_j),\quad j=A,B,
\end{eqnarray}
%%%%%%%%%%%%%%%%%%%%%%%%%%%%%%%%%%%%%%%%%%%%%%%%%%%%%%%%%%%%%%%%%%%%%%%%%%%%%%%%%%%%%%%%%%%%%%%%%%%%%%%%%%
and satisfying the completeness condition $\int_{z_j} \hat{\pi}_j(z_{j})dz_{j}= \hat{\openone}_j$.
Here $p_{j}(y_{j})$ is a normalized distribution of the parameter $y_{j}$, $\hat{\Pi}_j(y_{j})$ is a
normalized Gaussian state with CM $\gamma_{j}(y_{j})$ dependent on parameter $y_{j}$ and $z_{j}=(d_{j}^{T},y_{j}^{T})^{T}$.
%%%%%%%%%%%%%%%%%%%%%%%%%%%%%%%%%%%%%%%%%%%%%%%%%%%%%%%%%%%%%%%%%%%%%%%%%%%%%%%%%%%%%%%%%%%%%%%%%%%%%%%%
%\begin{equation}
%\frac{1}{2\pi} \int_{d_j,y_j} \hat{D}_j(d_j) \hat{\Pi}_j(y_{j}) \hat{D}_j^\dagger(d_j)d^{2}d_{j}dy_{j}=\hat{\openone}_j.
%\label{completness2}
%\end{equation}
%%%%%%%%%%%%%%%%%%%%%%%%%%%%%%%%%%%%%%%%%%%%%%%%%%%%%%%%%%%%%%%%%%%%%%%%%%%%%%%%%%%%%%%%%%%%%%%%%%%%%%%
Upon measuring the POVM (\ref{noncovariant}) on the Gaussian state $\hat{\rho}_{AB}$, one finds the
outcomes $z_{A}$ and $z_{B}$ to follow the distribution
${\cal P}(z_{A},z_{B})=p_{A}(y_{A})p_{B}(y_{B})P(d,y_{A},y_{B})$, where the distribution
$P(d,y_{A},y_{B})$ is obtained from Eq.~(\ref{Pd}) by replacing
$\gamma_{j}$ with $\gamma_{j}(y_{j})$. Denoting the classical mutual information of the
distribution ${\cal P}(z_{A},z_{B})$ and $P(d,y_{A},y_{B})$ as ${\cal I}[A(z_{A}):B(z_{B})]$ and
${\cal I}[A(y_{A}):B(y_{B})]$, respectively, one then has
%%%%%%%%%%%%%%%%%%%%%%%%%%%%%%%%%%%%%%%%%%%%%%%%%%%%%%%%%%%%%%%%%%%%%%%%%%%%%%%%%%%%%%%%%%%%%%%%%%%%%%%%%
\begin{eqnarray}\label{Iz}
{\cal I}[A(z_{A}):B(z_{B})]&=&\int{\cal I}[A(y_{A}):B(y_{B})]\mathop{\Pi}_{j=A,B}p(y_{j})dy_{j}.\nonumber\\
\end{eqnarray}
%%%%%%%%%%%%%%%%%%%%%%%%%%%%%%%%%%%%%%%%%%%%%%%%%%%%%%%%%%%%%%%%%%%%%%%%%%%%%%%%%%%%%%%%%%%%%%%%%%%%%%%%%
Hence it follows immediately that
%%%%%%%%%%%%%%%%%%%%%%%%%%%%%%%%%%%%%%%%%%%%%%%%%%%%%%%%%%%%%%%%%%%%%%%%%%%%%%%%%%%%%%%%%%%%%%%%%%%%%%%%%
\begin{eqnarray}\label{Inoncov}
{\cal I}_{c}^G(\hat{\rho}_{AB})&\leq&\mathop{\mbox{max}}_{\hat{\pi}_{A}(z_{A})\otimes\hat{\pi}_{B}(z_{B})}{\cal I}[A(z_{A}):B(z_{B})]\nonumber\\
&=&{\cal I}[A(y_{A}^{0}):B(y_{B}^{0})],
\end{eqnarray}
%%%%%%%%%%%%%%%%%%%%%%%%%%%%%%%%%%%%%%%%%%%%%%%%%%%%%%%%%%%%%%%%%%%%%%%%%%%%%%%%%%%%%%%%%%%%%%%%%%%%%%%%%
where we have to maximize over all Gaussian POVM elements $\hat{\pi}_{j}(z_{j})$ and
$y_{j}^{0}$, $j=A,B$ label the POVM elements $\hat{\Pi}_j(y_{j}^{0})$, which maximize ${\cal I}[A(y_{A}):B(y_{B})]$.
If we take these as seed elements of POVMs (\ref{POVM}), we construct local
covariant POVMs which give mutual information ${\cal I}[A(y_{A}^{0}):B(y_{B}^{0})]$ and therefore achieve the classical
mutual information ${\cal I}_{c}^G(\hat{\rho}_{AB})$.


\begin{thebibliography}{99}

\bibitem{nielsen}
M.~A. Nielsen and I.~L. Chuang, \emph{Quantum Computation and Quantum
  Information}\ (Cambridge University Press, Cambridge, 2000).

\bibitem{hororev} R. Horodecki,  P. Horodecki, M. Horodecki, and K. Horodecki, Rev. Mod. Phys. {\bf 81}, 865 (2009).

\bibitem{OZ} H. Ollivier and W. H. Zurek, Phys. Rev. Lett. {\bf 88}, 017901 (2001).
\bibitem{HV} L. Henderson and V. Vedral, J. Phys. A {\bf 34}, 6899 (2001).

\bibitem{ferraro}
A. Ferraro, L. Aolita, D. Cavalcanti, F. M. Cucchietti, and A. Acin
 Phys. Rev. A {\bf 81}, 052318 (2010).

\bibitem{piani}
M. Piani,
P. Horodecki, and R. Horodecki,
Phys. Rev. Lett. \textbf{100}, 090502 (2008).

\bibitem{dattaqc} A. Datta, S. T. Flammia, and C. M. Caves,
Phys. Rev. A \textbf{72} 042316 (2005);
A. Datta, and G. Vidal, Phys. Rev. A {\bf 75}, 042310 (2007);
A. Datta, A. Shaji, and C. M. Caves,
Phys. Rev. Lett \textbf{100}, 050502 (2008).

\bibitem{barbieri} B. P. Lanyon, M. Barbieri, M. P. Almeida, and A. G. White, Phys. Rev. Lett. {\bf 101}, 200501 (2008).

\bibitem{brasilearxiv}
K. Modi, M. Williamson, H. Cable, and V. Vedral, arXiv:1003.1174;
B. Eastin, arXiv:1006.4402; F. F. Fanchini, M. F. Cornelio, M. C. de Oliveira, A. O. Caldeira, arXiv:1006.2460; A. Brodutch and D. R. Terno, Phys. Rev. A {\bf 83}, 010301 (2011).

\bibitem{MID} S. Luo, Phys. Rev. A {\bf 77}, 022301 (2008).

\bibitem{moelmer} S. Wu, U. V. Poulsen, and K. M{\o}lmer, Phys. Rev. A {\bf 80}, 032319 (2009).

\bibitem{amid} D. Girolami, M. Paternostro, and G. Adesso, arXiv:1008.4136.

\bibitem{misure} J. Oppenheim, M. Horodecki, P. Horodecki, and R. Horodecki, Phys. Rev. Lett. {\bf 89}, 180402 (2002); M. Horodecki,
J. Oppenheim, and R. Horodecki, {\it ibid.} {\bf 89}, 240403 (2002) M. Piani, M. Christandl, C. E. Mora, and P. Horodecki, {\it ibid.} {\bf 102}, 250503 (2009); K. Modi, T. Paterek, W. Son, V. Vedral, and M. Williamson, {\it ibid.} {\bf 104}, 080501 (2010); B. Daki\'c, C. Brukner, and V. Vedral, {\it ibid.} {\bf 105}, 190502 (2010); S. Luo, and S. Fu, Phys. Rev. A {\bf 82}, 034302 (2010).

\bibitem{operdiscord} D. Cavalcanti, L. Aolita, S. Boixo, K. Modi, M. Piani, and A. Winter, arXiv:1008.3205; V. Madhok and A. Datta, arXiv:1008.4135.

\bibitem{ourreview} G. Adesso and F. Illuminati, J. Phys. A \textbf{40} 7821, (2007).

\bibitem{gaussexp} S. L. Braunstein and P. van Loock, Rev. Mod. Phys. {\bf 77}, 513
(2005); N.~Cerf, G.~Leuchs, and E.~S. Polzik (eds.), \emph{Quantum
Information with
  Continuous Variables of Atoms and Light}\ (Imperial College Press, London,
  2007).

\bibitem{AD_10} G. Adesso, and A. Datta, Phys. Rev. Lett. {\bf 105}, 030501 (2010).

\bibitem{Giorda_10} P. Giorda and M. G. A. Paris, Phys. Rev. Lett. {\bf 105}, 020503 (2010).

\bibitem{brasilesymm} J. Maziero, L. C. Celeri, and R. Serra, arXiv:1004.2082.

\bibitem{terhal} B. M. Terhal, M. Horodecki, D. W. Leung, and D. P.DiVincenzo,  J. Math. Phys. {\bf 43}, 4286 (2002); D. P. DiVincenzo, M. Horodecki, D. Leung, J. Smolin, and B. M. Terhal,  Phys. Rev. Lett. {\bf 92}, 067902 (2004).

\bibitem{Cerf05}
N. J. Cerf, O. Kr\"{u}ger, P. Navez, R. F. Werner, and M. M. Wolf,
Phys. Rev. Lett. \textbf{95}, 070501 (2005).

\bibitem{Mista06}
L. Mi\v{s}ta, Jr., Phys. Rev. A \textbf{73}, 032335 (2006).

\bibitem{Fiurasek_07} J. Fiur\'a\v{s}ek and L. Mi\v{s}ta, Jr., Phys. Rev. A {\bf 75}, 060302(R) (2007).

\bibitem{menicucci} N.~C. Menicucci, P.~{van Loock}, M.~Gu, C.~Weedbrook, T.~C. Ralph, and M.~A.
  Nielsen, Phys. Rev. Lett. \textbf{97}, 110501 (2005).

\bibitem{giedke03} G. Giedke, M. M. Wolf, O. Kr\"uger,
R. F. Werner, and J. I. Cirac, Phys. Rev. Lett. {\bf 91}, 107901
(2003).

\bibitem{geof} M. M. Wolf, G. Giedke, O. Kr\"uger,
R. F. Werner, and J. I. Cirac, Phys. Rev. A {\bf 69}, 052320 (2004).


\bibitem{D'Ariano_04} G. M. D'Ariano, P. Perinotti, and M. F. Sacchi, J. Opt. B: Quantum Semiclassical Opt.
{\bf 6}, 487 (2004).

\bibitem{Stratonovich_65} R. L. Stratonovich, Izv. Vyssh. Uchebn. Zaved., Radiofiz. {\bf 8}, 116 (1965) [Probl. Inf. Transm. {\bf 2}, 35 (1966)].

\bibitem{Adami_97} C. Adami and N. J. Cerf, Phys. Rev. A {\bf 56}, 3470 (1997).

\bibitem{Holevo_01} A. S. Holevo and R. F. Werner, Phys. Rev. A {\bf 63}, 032312 (2001).

\bibitem{Williamson_36} J. Williamson, Am. J. Math. {\bf 58}, 141 (1936).

\bibitem{extremal} J. Eisert, Ph.D. thesis, University of Potsdam, 2001; G. Vidal and R. F. Werner, Phys. Rev. A {\bf 65}, 032314 (2002);
G. Adesso, A. Serafini, and F. Illuminati, Phys. Rev. A \textbf{69}, 022309 (2004).

\bibitem{spinchains} T. Werlang, C. Trippe, G. A. P. Ribeiro, and G. Rigolin, Phys. Rev. Lett. {\bf 105}, 095702 (2010).

\bibitem{decoherence} L. Mazzola, J. Piilo, and S. Maniscalco, Phys. Rev. Lett. {\bf 104}, 200401 (2010); J.-S. Xu, X.-Y. Xu, C.-F. Li, C.-J. Zhang, X.-B. Zou, and G.-C. Guo, Nat. Commun. {\bf 1}, 7 (2010) and references therein.

%\bibitem{usediscord}  R. Dillenschneider, Phys. Rev. B {\bf 78}, 224413 (2008); C. A. Rodriguez-Rosario, K. Modi, A. Kuah, A. Shaji, and %E. C. G. Sudarshan, J. Phys. A: Math. Theor. {\bf 41}, 205301 (2008); A. Shabani and D. A. Lidar,
%Phys. Rev. Lett. {\bf 102}, 100402 (2009);  M. S. Sarandy, Phys. Rev. A {\bf 80}, 022108 (2009); T. Werlang, S. Souza, F. F. Fanchini, %and C. J. Villas-Boas, Phys Rev. A {\bf 80}, 024103 (2009); B. Bylicka, and D. Chruscinski,  Phys. Rev. A {\bf 81}, 062102 (2010);
%L. Mazzola, J. Piilo, and S. Maniscalco, Phys. Rev. Lett. {\bf 104}, 200401 (2010); L. C. Celeri, A. G. S. Landulfo, R. M. Serra, and G. %E. A. Matsas, Phys. Rev. A {\bf 81}, 062130 (2010); A. Brodutch, and D. R. Terno,  Phys. Rev. A {\bf 81}, 062103 (2010); M. D. Lang, and %C. M. Caves, Phys. Rev. Lett. {\bf 105}, 150501 (2010); T. Werlang, C. Trippe, G. A. P. Ribeiro, and G. Rigolin, Phys. Rev. Lett. {\bf %105}, 095702 (2010); F. F. Fanchini, L. K. Castelano, and A. O. Caldeira, New J. Phys. {\bf 12}, 073009 (2010); B. Wang, Z. Xu, Z. Chen, %M. Feng, Phys. Rev. A {\bf 81}, 014101 (2010); F. F. Fanchini, T. Werlang, C. A. Brasil, L. G. E. Arruda, and A. O. Caldeira, Phys. Rev. %A. {\bf 81}, 052107 (2010); D. O. Soares-Pinto, L. C. Celeri, R. Auccaise, F. F. Fanchini, E. R. deAzevedo, J. Maziero, T. J. Bonagamba, %and R. M. Serra, Phys. Rev. A {\bf 81}, 062118 (2010);
%J.-S. Xu, X.-Y. Xu, C.-F. Li, C.-J. Zhang, X.-B. Zou, and G.-C. Guo, Nat. Commun. {\bf 1}, 7 (2010); K. Bradler, M. M. Wilde, S. %Vinjanampathy, and D. B. Uskov, arXiv:0912.5112;
%A. Datta, arXiv:1003.5256; M. F. Cornelio, M. C. de Oliveira, and F. F. Fanchini, arXiv:1007.0228;
%A. Al Qasimi, and D. F. V. James, arXiv:1007.1814; F. Galve, G. L. Giorgi, and R. Zambrini, arXiv:1007.2174.

\bibitem{luoalber} S. Luo, Phys. Rev. A {\bf 77}, 042303 (2008); M. Ali {\it et al.}, Phys. Rev. A {\bf 81}, 042105 (2010).

\bibitem{Shannon_48} C. E. Shannon, Bell Syst. Tech. J. {\bf 27}, 623 (1948).

\bibitem{useMID} A. Datta and S. Gharibian, Phys. Rev. A {\bf 79}, 042325 (2009); A. Datta, Phys. Rev. A {\bf 80}, 052304 (2009); R. Srikanth, S. Banerjee, and C. M. Chandrashekar, Phys. Rev. A {\bf 81}, 062123 (2010); A. Auyuanet and L. Davidovich, Phys. Rev. A {\bf 82}, 032112 (2010).

\bibitem{bennet96} C. H. Bennett, D. P. DiVincenzo, J. Smolin, and W. K. Wootters,
Phys. Rev. A \textbf{54}, 3824 (1996).

\bibitem{ordering} G. Adesso and F. Illuminati, Phys. Rev. A {\bf  72},  032334 (2005).

\bibitem{Simon_00} R. Simon, Phys. Rev. Lett. {\bf 84}, 2726 (2000).

\bibitem{Perina_05} J. Pe\v{r}ina and J. K\v{r}epelka, J. Opt. B: Quantum Semiclassical Opt. {\bf 7}, 246 (2005).

\bibitem{Perina_91} J. Pe\v{r}ina, {\it Quantum Statistics of Linear and Nonlinear Optical Phenomena} (Kluwer, Dordrecht, 1991).

\bibitem{Dodonov} V. Dodonov, O. Manko and V. Manko, Phys. Rev. A \textbf{49}, 2993 (1994); {\it ibid.} \textbf{50}, 813 (1994).

\bibitem{Gelfand_57} I. M. Gelfand and A. M. Yaglom, Usp. Mat. Nauk {\bf 12}, 3 (1957).

\bibitem{Leonhardt_97} U. Leonhardt, {\it Measuring the Quantum
State of Light} (Cambridge University Press, Cambridge, 1997).

\bibitem{Kolmogorov_56} A. N. Kolmogorov, IRE Trans. Inf. Theory IT-{\bf 2}, 102 (1956).


\bibitem{Mista_08} L. Mi\v{s}ta, Jr., and J. Fiur\'{a}\v{s}ek, Phys. Rev. A {\bf 78}, 012359 (2008).


\bibitem{James} A. Al Qasimi and D. F. V. James, Phys. Rev. A {\bf 83}, 032101 (2011).

\end{thebibliography}
\end{document}